\documentclass[11pt]{article}
\pdfoutput=1
\usepackage[utf8]{inputenc}
\usepackage{comment, graphicx, url, xr, zref, algpseudocode, setspace, ulem}
\usepackage{algorithm, amsthm, algorithmicx, authblk, bm, color, graphicx, amssymb, amsmath, amsfonts}

\newtheorem{lem}[]{Lemma}
\theoremstyle{definition}
\newtheorem{ex}[]{Example}
\newtheorem{re}[]{Remark}

\newcommand{\numeq}[1]{\begin{align}#1\end{align}}
\newcommand{\eq}[1]{\begin{align*}#1\end{align*}}
\newcommand{\E}{\mathbb{E}}

\newcommand{\Var}{\textnormal{Var}}

\date{\today}

\title{Bayesian Uncertainty Directed Trial Designs}
 
\author{
Steffen Ventz$^{1,2,\star}$\footnote{steffen@jimmy.harvard.edu}, 
Matteo Cellamare$^{1,2,\star}$\footnote{matteoc@jimmy.harvard.edu},
Sergio Bacallado$^{3}$\footnote{sb2116@cam.ac.uk},
and 
Lorenzo Trippa$^{1,2}$\footnote{ltrippa@jimmy.harvard.edu}\\
$^{1}${\small Dana-Farber Cancer Institute, US},
$^{2}${\small Harvard T.H. Chan School of Public Health, US},\\
$^{3}${\small University of Cambridge, UK}\\
$\star${\small co-first authorship}
}

\begin{document}

\maketitle

\begin{abstract}
\noindent {\small 
 Most Bayesian response-adaptive designs  unbalance  randomization rates  towards the most promising arms with  the  goal  of  increasing  the number of positive treatment outcomes during the study, even though  the  primary  aim of the trial  is different. We discuss Bayesian uncertainty directed designs (BUD),  a class of  Bayesian  designs in which the investigator specifies an information measure  tailored to the experiment. All decisions during  the  trial  are  selected to   optimize   the available information at the end of the study. The approach  can be applied to several   designs, ranging from early stage multi-arm trials to  biomarker-driven and multi-endpoint studies.  We discuss the asymptotic  limit of the  patient allocation proportion to treatments, and illustrate  the finite-sample operating characteristics   of BUD designs through  examples, including  multi-arm trials,  biomarker-stratified  trials, and trials with multiple co-primary endpoints.}
 
{\bf Keywords:} Multi-arm clinical trials, Information theory, Decision theory, 
Response-adaptive designs.
\end{abstract}

\section{Introduction}


We discuss a  class  of Bayesian adaptive designs   for randomized clinical trials  that seek to maximize the acquisition of information on the effectiveness of new  experimental treatments. 
We 
call this class of designs Bayesian uncertainty directed  (BUD) designs. 
For a BUD design, the  investigator 
specifies a Bayesian model  that  will  be continuously  updated  during  the  trial     and 
 a metric  to quantify the accumulated information on experimental treatments.  
This information measure is a summary  of the posterior  distribution, 
and   all decisions during  the  trial  are  selected to approximately  optimize   the available information  on  experimental  treatments  at the end of the study.
We illustrate  the approach through  examples, including   controlled multi-arm trials, 
biomarker-defined subgroup trials  \cite{Ventz2017}, and trials with multiple co-primary endpoints  \cite{Offen2007}.

Adaptive designs for clinical experiments use
 the accumulating data during  the study to modify  
characteristics of the ongoing trial  \cite{berry2010bayesian, thall2010bayesian}.
For instance patient eligibility  can be modified during the study, 
the overall sample size may be re-estimated at interim analyses, 
or  randomization probabilities  can  be  unbalanced  towards the most promising arms  \cite{thall2007practical, Lee2010, trippa2012bayesian, Ventz2017}.
Adaptive designs  have been applied  in  several  settings,  ranging from dose finding  studies  to  Phase III trials   \cite{Schoffski2004,  barker2009spy, kim2011battle, Cellamare2016, Cellamare2}.
These  designs   have been developed to save resources, to protect patients from ineffective treatments, and to gain efficiency in the   development of  new treatments.

For most  clinical studies  one   can  consider  multiple    
designs, including  adaptive randomization   \cite{berry2010bayesian}.
    Practitioners  can  compare and select from competing candidate methodologies.
Selecting   according   to  interpretable  performance criteria is  desirable  \cite{Ventz2015, Ventz2017b},
and it  forces  the   investigator to explicitly specify  the primary  aims of the  trial.     
The Bayesian decision-theoretic   framework    
identifies  the design that maximizes the performance criteria in expectation. 
The approach   enables the use of  utility functions $u$ to represent  investigator  preferences and 
  it incorporates {\it  a priori}  information into the design
 through  Bayesian modeling     \cite{Berger1985}. 
 The  utility $u=u(Y,\theta, d)$ is a random  
 function  of  the unknown parameters $\theta$, for instance  response probabilities that one seeks  to  estimate  during  the  trial,
and  of the data $Y$ collected during the study  using the  design $d$.
 Bayesian  designs maximize the utility in expectation.

Bandit  problems have been  studied extensively in the context of clinical trials  \cite{Berry1985, berry1988one, lewis1994group, Cheng2007a, zhang2015optimal, Ventz_2018_add}. 
A multi-arm Bandit problem   describes a sequential decision process.  
At each stage of the process, a decision maker chooses to ``play'' one of  $K$ arms, 
this arm     
then generates a random outcome and a corresponding payoff. 
In a clinical trial, the   arms typically correspond to $K$  treatments,
and the outcome summarizes  the   response of a patient to the assigned  treatment.
In a Bandit problem the decision maker seeks to select arms during the trial to maximize the  total  payoff of the sequential experiment  \cite{Berry1985}. 
This class of decision-theoretic
 designs can be used to maximize an explicit utility in expectation,    
for instance  the expected number of responders at the end of the trial.
The  solution of a Bandit problem  is  
 the sequence of decisions  that maximizes the expected payoff,  and  
  in  some  cases it  can  be computed  with  backward induction (BI)  \cite{Berger1985}.

It has  been  well  documented  that BI  is   often  computationally  prohibitive   \cite{Berger1985}, and several approximation  strategies have been suggested. 
For instance, 
Carlin et al. \cite{carlin1998approaches} 
introduced a forward sampling 
algorithm.
%
%
%
Alternative  algorithms  \cite{kadane2002hybrid, brockwell2003gridding, muller2007simulation} approximate  Bandit  solutions  
by combining forward sampling with backward induction.   
%
%
  M-steps  myopic  look-ahead procedures, which  select at each state  the action that   maximizes the expected utility  at  completion of  a  fictitious   
 M-stages horizon, 
 have been   discussed in 
 \cite{bubeck2012regret}. 
These procedures can    be extended to allow  for  randomized  actions   \cite{russo2014learning}.  Indeed randomization  is  often  a  requirement  in  clinical  trials.
    In  these   cases,  for  each patient 
 an arm $a$ is randomly  assigned, with  randomization probabilities  chosen   within a  fixed  and pre-specified    subset  of  the  simplex,
   to  optimize  the   payoff  at   M-stages.
   %

\begin{figure}[h]
\begin{center} 
\includegraphics [width=17cm, height=9cm]  {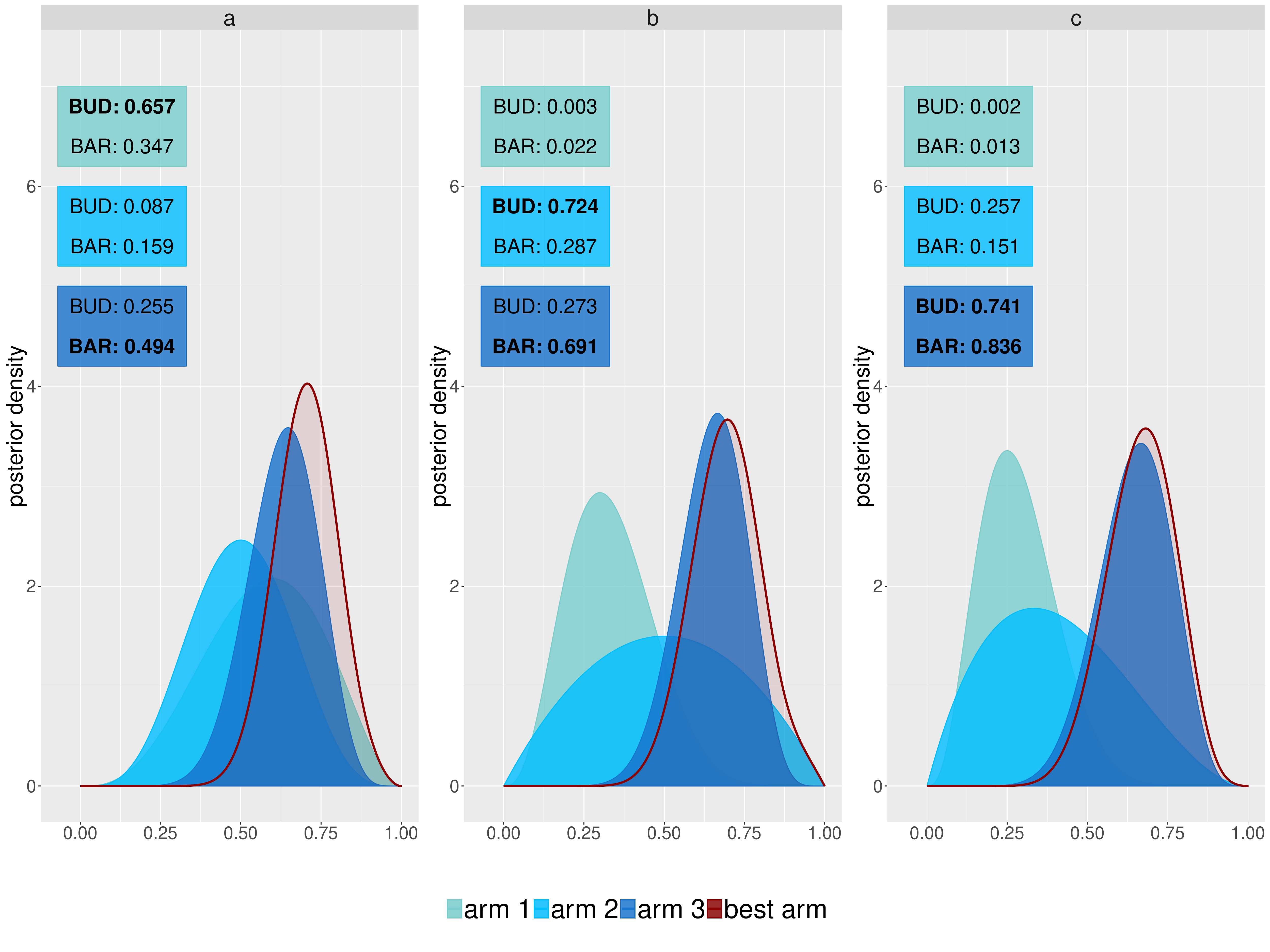}
\caption{
Comparison of a Bayesian uncertainty directed  (BUD) design and 
Bayesian adaptive randomization (BAR).  We  consider an early stage  3-arm trial without control,  with  the primary goal of  selecting  the  best  experimental arm  $a^\star = \arg \max_{a=1,2,3} \theta_a $. The  total sample  size is equal to $T=100$.
BAR  defines  randomization probabilities to   treatments $a=1,2,3$ that  mirror  the  posterior  distribution of  $a^*$  \cite{thompson1933likelihood}.
The BUD design randomizes  patients  with  the goal of   approximately minimizing  the posterior entropy of   $a^\star$. Panels (a), (b) and  (c) show three different simulations  of  the  3-arm trial after  the enrollment  of  50  patients. The blue curves show the posterior densities   of the  response probabilities $\theta_a$. The red curves indicate  the posterior distribution  of the maximum $\theta_{a^\star}=\max_{a=1,2,3} \theta_a$.  In each panel  we  also  indicate  for  BAR  and BUD designs the randomization probabilities  of  the next patient $t=51$ that  will  be  enrolled in the trial.
}\label{Fig:BEST}
\end{center}
\end{figure}

Popular Bayesian  designs, including Bayesian adaptive randomization (BAR)  \cite{thall2007practical},  unbalance  randomization probabilities  towards the most  promising arms.
  These designs appear   consistent  with  the  goal  of  maximizing  the number of positive treatment outcomes during the study,
although   the  primary  aim of the trial  could   be different. 
Here we focus on uncertainty directed designs for  clinical trials.   Figure \ref{Fig:BEST}  illustrates  relevant  differences  between the   randomization probabilities  
of  BUD  and BAR  designs.  In this  figure  each panel  summarizes  the  posterior  distribution  for a  3-arm  trial  (without  control)   
with the primary goal   of  selecting  the most effective treatment.
  
BUD designs use  
information measures that are   convex     functionals of the posterior  distribution  $p(\theta \mid   Y)$.  
   Examples will include    the    entropy of the posterior probability 
of a positive treatment effect  $H( I(\gamma_a(\theta) >0 ) \mid Y)$  for an experimental arm $a$, 
and  the  posterior variance of treatment effects  $\Var(  \gamma_a(\theta) \mid  Y  )$.    
The use of  a   convex     measure ensures  ---  by straightforward arguments  based on  Jensen's inequality  ---  that,  on average,   information  increases with every  additional   observation. 
In a BUD design decisions during the trial are selected to  maximize    these   information increments. 
Using simulations, 
we show 
that  in multi-arm studies,
the operating characteristics  of    BUD designs  are
 comparable to   decision-theoretic designs  derived through BI  when the corresponding utility function $u$ and    information  metric coincide.

The paper is structured as follows. 
Section \ref{GeneralProposal} introduces BUD designs and  
illustrates  their asymptotic behavior   
with two specific examples.
%
%
Section \ref{Exa1} examines the operating characteristics of BUD designs for multi-arm trials. 
Section \ref{Sec:SubgroupTrial} discusses BUD designs for biomarker-defined subpopulation trials.
The aim here is to match  targeted treatments with  subgroups of patients that benefit from them.
%
%
%
%
%
Section \ref{Sec:CoPrimary} deals with an application of the BUD approach to a multi-arm trial with several co-primary endpoints  \cite{Offen2007}.  
We conclude the paper with a discussion in Section \ref{Discussion}.  
Details on the computational implementation of the methods proposed and  
R   functions  are provided  in the  supplementary material.   

%
%
%




\section{Bayesian Uncertainty Directed Designs}\label{GeneralProposal}
%
%
In the  decision-theoretic framework for sequential experiments the investigator specifies a probability model for    unknown parameters  and observations during    the trial.
At  pre-planned stages,  the experiment can be modified by selecting 
 actions, which 
may correspond  for example to  the decision to  close the  enrollment  for a  patient  subgroup or  to modify the randomization probabilities.

 The decision maker selects, at stages $t=1, \dots, T$,  an action $A_t $ from a set 
 $\mathcal A$ in a sequential manner.  
Each action $A_t$, in turn, generates a random  outcome  
$Y_{t}  \sim p(Y_t\mid  A_t, \theta)$  with value in $\mathcal Y$.  
We use a prior probability  
for the  parameters $\theta \in \Theta$ of  the  outcome distribution.  
The action taken at time $t+1$ is a random variable $A_{t+1}$, whose distribution depends on the history of previous actions and 
outcomes $\Sigma_t= \{ (A_\ell, Y_\ell), \ell \leq t \}$. We  write $A_{t+1}=d(\Sigma_t)$ to make this dependence explicit.
In different  words,  previous actions and 
outcomes $\Sigma_t$ are  translated    either  into the selection of  a   point  in $\mathcal{A}$,  for  instance  the selection of  a  specific  arm,
or  into a  distribution over  $\mathcal{A}$,   for example   the  randomization probabilities  for   patients that will be enrolled.

Several authors proposed and justified  multi-stage designs 
using utility functions \cite{jung2004admissible}.  
In a two-stage single-arm trial with  binary endpoints with  $n_t$ patients  for stages $t=1,2$,
{   the 
  actions $(A_2, A_3) \in \{0,1\}^2$ 
represent the decisions 
(i) to continue the trial after the first   interim analysis  if  the number  of  responses  is  sufficiently  promising    $A_2 = d(\Sigma_1) = I(Y_1 > y_1)$, and 
subsequently (ii)  to recommend the treatment for a confirmatory trial   if more than $y_2> 0$  responses have been observed during the trial,  $A_3 = d(\Sigma_2) = I(Y_1 > y_1)\times I(Y_1+Y_2>y_2)$.  
} 
In the example $Y_t$ is a binomial random variable with size $n_t$ and probability $\theta$. 

For a given experiment, 
we let $\mathcal D$ denote the set of  decision functions that map the data into actions.
For the two-stage design  $\mathcal D$ can be identified by varying the thresholds
$ (y_1, y_2) \in \{ 0, \dots, n_1\} \times \{ 0, \dots, n_1+n_2\}$.
%
%
%
%
%
Throughout  the article $u(d,Y,\theta)$  indicates   
the utility function.
 It is a random quantity, which is 
a function of the parameter $\theta$ and the data $Y$ generated  under design $d \in \mathcal D$. 
A rational investigator selects 
the  design $d$ from $\mathcal D$ that maximizes the expected utility
$U(d) = \mathbb E [ u(d,Y,\theta) ]$
of the experiment  \cite{Berger1985}, 
\begin{align}\label{d:OB}
d^\star = \mathop{\arg \max}_{d \in \mathcal D} U(d).
\end{align}
In some cases  $ d^\star$ can be computed exactly using backward induction (BI)  \cite{Berger1985}, but this   is often infeasible. 
{   The   two-arm  bandit  problem  with  binary  outcomes, 
as  described  in   \cite{Berry1985, Cheng2007a}, is a  good  example  
to describe  how  quickly the computational burden of 
the BI algorithm  increases with  the  sample  size.
For a  trial with $50$ observations,  
the BI  algorithm requires $\binom{54}{4}  =  316,251$  operations  dedicated to
each possible configuration  of  the  sufficient  statistics, these  are the
combinations    of  positive  and  negative  outcomes for each arm after $t=0,1, \cdots, T$ assignments.
The number of operations  increases quickly with the sample size,  for  instance,  by  doubling  the  sample  size the  number  of operations increases to $\binom{104} {4} =   4,598,126$.  
Strictly related  considerations   have been  discussed for biomarker  trials  in  
 \cite{zhang2015optimal}.       }

We focus  therefore on a myopic  approximation  of $ d^\star$   \cite{Berger1985},  and proceed in three steps:

\textsf{Step (1) -  Action space:} 
We first specify   the set of  actions  $\mathcal A$
 that can be selected at each of the $T$ stages of the experiment. 
For  example,  in a 
 controlled two-arm trial, which  randomizes $n$  patients  during   each stage $t$, the action $A_t$ is the number of patients assigned to  the experimental arm.
  Hence, $
 \mathcal A = \{0,1,\dots,n\}$. The randomization  probability  $r_t\in (0,1)$ 
  toward  the experimental  arm is a function of the history of actions and outcomes $\Sigma_{t-1}$, 
  and $A_{t}\mid \Sigma_{t-1} \sim\text{Binomial}(n,r_t)$. 

\textsf{Step (2) - Information metric:} 
Next  we quantify the acquired information  through the accumulated data  $\Sigma_t$ until    stage  $t$.
Our utility  function $u(\cdot)$  will   quantify the information accrued by the experiment. 
Large values of $u(\Sigma_t)$ correspond to  low  uncertainty levels.
We  use   functions $\tilde{u}$
 that translate 
  the posterior $p(\theta \in \cdot  \mid \Sigma_t)$ into  utilities,
  i.e. $u(\Sigma_t) = \tilde{u}( p( \theta \in \cdot \mid \Sigma_t) )$.
The utility functions  $u$    can  have  negative values.  
In the  next paragraphs we   provide some   examples:
\begin{itemize}

\item [(i)] {\it Multi-arm trials:}  
Consider  a multi-arm  study with binary endpoints
and   primary  aim  of  providing  accurate   
estimates of the treatment effects $\gamma_a$  for    the effective experimental  treatments   $ a $ with 
 $\gamma_a>0$. 
%
%
Here $\gamma_a = \theta_a - \theta_0$  is  the  difference  between the response probability $\theta_a$ for treatment $a$ and the control arm.
%
We can define  
\numeq{ \label{multi-arm trial metric}
u(\Sigma_t)=  \sum_{a=1}^K \Big( v_a - 
\mbox{Var}( \gamma_a \times I(\gamma_a>0) \mid  \Sigma_t ) \Big)
}
to measure  
information up to stage $t$, 
with $v_a=\mbox{Var}(\gamma_a \times I(\gamma_a>0))$ denoting the prior variance.

\item [(ii)] {\it Biomarker-stratified trial:} 
We 
test  $K$  
 treatments for patients with and without a  genomic alteration. 
The trial  will  measure    
binary outcomes 
 with parameters  $\theta_{x,a}$ for subgroups $x=0,1$  and treatments $a=0,\dots,K$. 
The primary aim is to test the presence of    effects  within  subgroups  $E_{x,a} = I(\theta_{x,a}>\theta_{x,0})$ 
 and in the overall population $E_{a}=I(\theta_{a}>\theta_{0})$.  Here
    $\theta_{a} = \beta \theta_{1,a} + (1-\beta)\theta_{0,a}$ and $\beta \in [0,1]$ is the prevalence of the biomarker.
Let   $H[ p(X) ]= -\mathbb E[ \log p (X) ] $  indicate the  
entropy of a  random variable $X$.  
We   can  use  a summary $u$  that  weights the  entropy  values  associated to     interpretable  posterior probabilities, 
$$ u(\Sigma_t)= -\sum_{a=1}^{ K } \bigg \{   H[ p( E_a \mid  \Sigma_t)] 
+w \times \Big( H[ p(E_{1, a}  \mid  \Sigma_t)] + 
H[ p( E_{0,a}  \mid  \Sigma_t) ] \Big) \bigg \}, \text{  with } w\ge 0.$$

\item [(iii)] { \it Dose-finding trial:}   
{ 
We select  one of $K$ candidate dose levels  $\mathcal A = \{1,2,\ldots, K\}$  
using  binary efficacy and  toxicity outcomes. 
We let $\theta_{E,a}$ and $\theta_{T,a}$ denote the probabilities of  response and toxicity  at dose level $a$.
For each  dose level $a$ a score    weights  efficacy   
$\theta_{E,a}$ and     toxicity $\theta_{T,a}$, say $\mathcal S_a(\theta) = w \theta_{E,a} + (1-w)(1-\theta_{T,a})$ with $0 \leq w \leq 1$ and
 dose level { $\displaystyle A^\star = \underset{ a }{\operatorname{arg\,max}} 
 ~   \mathcal S_a(\theta)$}  has  the highest  score. 
In  Bayesian modeling  $(\theta_{E,a}, \theta_{T,a})_a$,  as well as $\mathcal S_a(\theta)$ and   $A^\star$, are  random variables. 
The posterior distribution  $\displaystyle p( A^\star = a | \Sigma_t) = p\big( \cap_{a'} \{\mathcal S_a(\theta) \geq \mathcal S_{a'}(\theta) \}  | \Sigma_t\big), a=1, \cdots,K,$  changes over time as more information becomes available.
We can  use the  (negative) entropy of the posterior and specify
$u(\Sigma_t) =   \sum_a p(A^\star =a | \Sigma_t) \log p( A^\star =a  | \Sigma_t).$
}

\end{itemize}
%
%
For  each example we only mention one  information  metric   tailored to  the  aim of the trial.
Several alternative measures $u$   could be used.
The unifying element is the use   of  functionals  of the posterior to quantify information.

\textsf{Step (3) - Myopic  approximation:}   
%
As we   mentioned  the information metric 
$\tilde{u}$ 
is  specified by  a   convex    functional  over the convex  space  of  distributions  on $\Theta$.  In different  words 
$$ \tilde u( w \times p_1 + (1-w) \times p_2)    \leq  w  \times \tilde  u( p_1) + (1-w) \times \tilde u( p_2)$$
  for every  pair  of probability  measures  { $p_1$ and $p_2$, }
  when  $w \in  [0,1]$.
%
%
Let  $u(\Sigma_t)= \widetilde u (p(\theta \in \cdot \mid \Sigma_t))$ be  the current value of 
the  information function. 
By 
Jensen's inequality,  given the action selected at the next step $A_{t+1}=a$, the information on average  increases   
%
%
%
%
%
\begin{align}\label{average:increase}
\mathbb E[ u( \Sigma_{t+1}) \mid  \Sigma_{t}, A_{t+1}=a  ]   
&= \mathbb E[ \tilde{u}(p(  \cdot \mid  \Sigma_{t+1} ) ) \mid  \Sigma_{t}, A_{t+1}=a ]  \nonumber \\
& \ge 
\tilde{u}(\mathbb E[p(  \cdot \mid  \Sigma_{t+1} ) \mid  \Sigma_{t}, A_{t+1}=a] ) =   u( \Sigma_t ).
\end{align}
 It is desirable for $\tilde u$ to satisfy this inequality, as any additional observation should tend to reduce uncertainty. 
%
 
BUD designs   select actions that  generate  large    information increments. 
The myopic decision rule $\tilde{d} \in \mathcal D$ selects at each stage the  
 action that  maximizes the gain of information  
 $A_{t+1} = \tilde{d}(\Sigma_{t} ) = \mathop{\arg \max}_{a \in \mathcal A }    
\Delta_{t}( a ),$ where  $\Delta_{t}( a ) = \mathbb E [ u( \Sigma_{t+1} )  \mid  A_{t+1}=a, \Sigma_t]-u(\Sigma_t)\ge 0.$  
In most clinical trials, non-randomized 
 policies like the myopic decision rule described above,  
are inappropriate  \cite{berry1997optimal}.
For  this  reason we  use the BUD design $d_{BUD}
$,  which  is  a  
randomized  version of  the myopic design. 
%
The  design  translates   $ \Sigma_t$  into  a  distribution  
on  $ \mathcal A$.   In  particular,   if  $A_{t+1}$   is    the treatment  assigned  to  the  next  patient, then
\begin{align}\label{BUD::General}
p( A_{t+1} = a \mid  \Sigma_t ) \propto \Delta_{t}(a)^{h(t)}.
\end{align}
Here $h(\cdot)$ is a non negative    function. 
With large values of $h(t)$ the the BUD design $d_{BUD}$  and the myopic design $\tilde{d}$   become   nearly identical.  
%
On the other  extreme,  with $h(t)=0$,  the  randomization probabilities  become  identical  across arms.
%
%
%
%
%
%
The choice of $h(\cdot)$
has the  goals  of 
(i) satisfying  the requirement to randomize patients 
and (ii)  the  approximate  optimization of     utility  criteria $u$. 


\subsection{Asymptotic Allocation Proportions in BUD Designs}\label{Sec:Asymptotic}
Our goal  is  the study of BUD designs with realistic sample sizes
(Sections \ref{Exa1}--\ref{Sec:CoPrimary}). 
 However, understanding  the asymptotic behavior of BUD policies is useful to interpret simulation results and to ascertain 
 if  the proportions of patients  allocated to different arms converges to a nearly  optimal limit.
 Here we discuss with examples asymptotic characteristics  of BUD designs. 
 For simplicity we  consider a constant parameter $h(\cdot)=h\geq 0$
  in (\ref{BUD::General}). 

Lemma \ref{asymptotics::lemma} is a technical result that  we will later use  to derive  asymptotic allocation proportions in two examples. 
A proof following  stochastic approximation arguments can be found in the appendix. 
Let $ \widehat p_{a,t}$  be  the proportion of samples allocated to arm $a$ by time $t$. 

\begin{lem}\label{asymptotics::lemma}
Consider the allocation of patients to arms $\mathcal A=\{0,1\}$ with a BUD design.
%
%
Define $F_t=-\widehat p_{0,t}+\Delta_{t}^h(0)/(\Delta_{t}^h(0)+\Delta_{t}^h(1))$. 
Suppose that, on a set of probability 1,  for any $\varepsilon>0$ there is a random time $T$, 
and number $c>0$ such that
 $F_t<-c$ wherever $ \widehat p_{0,t}>\rho_0+\varepsilon$, and $F_t>c$ wherever $ \widehat p_{0,t}<\rho_0-\varepsilon$  for all $t>T$. 
 Then $ \widehat p_{0,t}\to \rho_0$  a.s.
\end{lem}

\begin{re} \label{multiarm::remark}
Before applying 
this result to examples, we observe that for  some  utility criteria $u$
it is straightforward to extend the lemma to experiments with multiple arms $\mathcal A=\{0, \dots, K\}$. 
In particular, when the information metric $u(\Sigma_t)$ is an additive functional of the posterior of each parameter $\theta_a$,  $a\in\mathcal A$, for example $u(\Sigma_t) = - \sum_a \Var(\theta_a\mid  \Sigma_t)$, 
the information gain $\Delta_t(a)$ 
depends only on the prediction of  outcomes   from  a  single arm $a$. 
For any pair of arms $(a_1,a_2)$, the subsequence of samples  assigned to these two arms   is equivalent  to  a two-arm BUD design. Therefore, 
if the conditions of Lemma \ref{asymptotics::lemma} hold,
\eq{
\frac{ \widehat p_{a_1,t}}{ \widehat p_{a_1,t}+ \widehat p_{a_2,t}} \stackrel{\text{a.s.}}{\to}  \tilde\rho_{a_1,a_2}
}
for every $a_1,a_2\in \mathcal A$. 
Then, 
the allocation proportions $( \widehat p_{0,t},\dots, \widehat p_{K,t})$ converge to a limit 
$\rho = (\rho_0,\dots,\rho_K)$,  which is the unique solution to the linear system
\eq{
\sum_{a=0}^K \rho_a &= 1\;, \qquad \rho_{a_1} = \tilde \rho_{a_1,a_2} (\rho_{a_1}+\rho_{a_2}) \quad\text{for all } \{a_1,a_2\}\subset\{0,\dots,K\}.
}
\end{re}

\begin{ex}{\bf Multi-arm trial with normal outcomes.}
Assume the outcome $Y_t \mid A_t=a  \sim N( \theta_a, \sigma^2_a )$ is  normal with unknown mean $\theta_{a}$ and known variance $\sigma^2_a.$
We use independent $N(0,\sigma^2)$ prior  distributions for $\theta_a$, $a\in\mathcal A=\{0,\dots,K\}$ and 
show that in a BUD design,  driven by the information measure 
\begin{align*}
u(\Sigma_t)= - \sum_{a=0}^K \mbox{Var}( \theta_a \mid \Sigma_t)= - \sum_{a=0}^K
\dfrac{1}{\sigma^{-2}+t \widehat p_{a,t} \times \sigma_a^{-2} },
\end{align*}
%
%
  the arm-specific sample size proportions 
 converge a.s. to
 \numeq{ \label{limit}
 \rho_a= \frac{\sigma_a^{\frac{2h}{1+2h}}}{\sum_{\ell=0}^K \sigma_\ell^{\frac{2h}{1+2h}}} \quad\text{for all } a=0,\dots,K.
 }
 For large $h$ the BUD design becomes  nearly identical to the  myopic design,  
and  the limit  (\ref{limit})   coincides   with the  Neyman allocation  $\rho_a \propto \sigma_a$
\cite{jennison1999group}.

 To show  the convergence  to (\ref{limit}), we note that the information gain $ \Delta_{t}(a)$ is equal to
\begin{align}\label{ApproximateAllocation}
 \Delta_{t}(a)
 & = \dfrac{1}{ \sigma^{-2}+ t \widehat p_{a,t}\sigma_a^{-2}} -  
 \dfrac{1}{ \sigma^{-2}+(t \widehat p_{a,t}+1) \sigma_a^{-2}}  \nonumber \\
 & \qquad =   \dfrac{ \sigma^2_a }{ (\sigma^2_a /\sigma^2  +t \widehat p_{a,t}  )(\sigma^2_a /\sigma^2 +1 +t \widehat p_{a,t})   }
  =  \dfrac{ \sigma^2_a }{  t^2 \widehat p_{a,t}^2 } 
 + \mathcal O(  ( \widehat  p_{a,t}  t)^{-3}).
\end{align}
Using (\ref{ApproximateAllocation}), we can 
write
\numeq{\label{FT}
F_t 
=-
\widehat p_{0,t}+ \frac{ \Delta_{t}^h(0) }{ \Delta_{t}^h(0)+\Delta_{t}^h(1) } 
= 
 - \widehat p_{0,t}+
\frac{ \Big( \widehat p_{0,t}^{-2} \sigma_0^{2} + \mathcal O(t^{-1}\widehat p_{0,t}^{-3}) \Big)^h}
{ 
    \Big( \widehat p_{0,t}^{-2} \sigma_0^{2} + \mathcal O(t^{-1}\widehat p_{0,t}^{-3}) \Big)^h
 + \Big( \widehat p_{1,t}^{-2} \sigma_1^{2} + \mathcal O(t^{-1}\widehat p_{1,t}^{-3}) \Big)^h
 }.
}
Consider a similar sequence
\numeq{\label{2FT2}
\tilde F_t = - \widehat p_{0,t}+\frac{ \widehat p_{0,t}^{-2h}\sigma_0^{2h}}{ \widehat p_{0,t}^{-2h}\sigma_0^{2h}+(1- \widehat p_{0,t})^{-2h}\sigma_1^{2h}}.
}

Note that $\tilde F_t$ is strictly decreasing in $\widehat p_{0,t}\in[0,1]$ with a zero at $\rho_0$,
 which implies 
that for any $\varepsilon>0$ there is a $c>0$ satisfying  $\tilde F_t<-c$ wherever $ \widehat p_{0,t}>\rho_0+\varepsilon$, and $\tilde F_t>c$ wherever $ \widehat p_{0,t}<\rho_0-\varepsilon$.  These are the assumptions  of Lemma  \ref{asymptotics::lemma}.

We need  to  show that $F_t-\tilde F_t \to 0$ as $t\to \infty$ to  apply  the lemma   and  derive  (\ref{limit}). 
First,
   consider  the $F_t$ subsequence  when   
   $t^{-1/3+\delta}< \widehat p_{0,t} < 1- t^{-1/3+\delta}$ for some $\delta\in(0,1/3)$.
In this  case  $\widehat p_{0,t}$ does not approach zero or one  too quickly,   and 
 the   difference  between  (\ref{FT})  and  (\ref{2FT2}) 
vanishes  for large $t$.
 Second, when $\widehat p_{0,t}\leq t^{-1/3+\delta}$, it can be verified that $\Delta_t(1)/\Delta_t(0) \to 0$ as $t\to\infty$, and that $F_t\to 1$. 
 Conversely,  when $\widehat p_{0,t}>1-t^{-1/3+\delta}$, $F_t\to -1$. 
We can therefore apply the lemma to conclude that $ \widehat p_{0,t} \to \rho_0$. For a multi-arm  BUD  design, $K>1$,  it is sufficient to  apply Remark \ref{multiarm::remark}.

\end{ex}

\begin{ex}{ \bf Multi-arm trial with binary outcomes.}
Now consider the case where $Y_t$ is binary with $p(Y_t=1\mid A_t=a)=\theta_{a}$, and a prior  $ \theta_{a}\sim \mbox{Beta}(\alpha,\beta)$ for $a\in\mathcal A$. 
We consider the same information metric of the previous example and prove that the allocation proportions converge with probability 1 to the limit in (\ref{limit}) with  $\sigma^2_a = \theta_a(1-\theta_a)$.
In this example, the information gain $\Delta_t(a)$ has a closed-form expression which  simplifies  the derivation of the asymptotic result. The proof follows the same arguments of the previous example, and it is deferred to the appendix.
\end{ex}
%
%
%
{  
 The examples above are by no means exhaustive, but their simplicity permits a self-contained analysis,  which    points to     mathematical techniques.  
An important  case not   covered  here  is that of  BUD designs in which the utility function $u$ is not  additive on the treatment arms.
The information metric (\ref{multi-arm trial metric}) in Section \ref{GeneralProposal} is an example 
where Lemma \ref{asymptotics::lemma} is  not  directly  applicable. 
Such information metics are of practical relevance, for example  
when some of the experimental agents are potentially 
significantly inferior to the standard of care. 
} 
We believe  that  the  extensive  literature  on stochastic approximation can be used 
for  additional  asymptotic analyses  of  BUD  designs, including  non additive  utilities $u$. 
        The monograph   \cite{pemantle2007}  provides a useful  guide. The general strategy would consist of writing  the vector of allocation proportions $(\widehat {\bm p}_{0:K,t})_{t\geq 0}$ as a stochastic approximation process; namely,  one would express the increments 
$   \widehat {\bm p}_{0:K,t+1} - \widehat {\bm p}_{0:K,t} 
= \gamma_t(F( \widehat {\bm p}_{0:K,t} ) + \epsilon_{t+1}+ \mathcal O(t^{-1} ))$ for some sequences $(\epsilon_t)_{t\geq 0}$ and $(\gamma_t)_{t\geq 0}$ with  $\mathbb E[\epsilon_{t+1}\mid \Sigma_t]=0$, $\sum_t \gamma_t=\infty$, and  $\sum_t \gamma_t/t<\infty$. 
Then the  arguments  used to  prove   Lemma \ref{asymptotics::lemma}  
can be adapted to establish  that $\{\widehat {\bm p}_{0:K,t}\}_{t\in \mathbb N}$   approaches the solution of a differential equation.

\section{BUD  Designs for Phase II  Multi-Arm Trials}\label{Exa1}
We  discuss BUD designs for two multi-arm  Phase II  trials with  distinct aims: 
(i) to  identify  all    experimental  arms with  a  positive  treatment effect   \cite{Wason2012},  and     
(ii)  to select the  treatment  with the  most   favorable   outcome  distribution   \cite{stallard2008group}.

\subsection{Estimation of  Treatment  Effects  in a  Controlled  Multi-Arm Study }\label{Sec:Multi:Arm1}
We first considered a  study with $K$ experimental  arms compared to a control therapy.
For each  patient $t$,  action $A_t= a$, with $a   \in \mathcal A= \{0, \dots, K\}$, 
indicates the   assignment of patient $t$ to arm $a$, where $a=0$ denotes the control arm. 
The  primary  outcome  is  the   binary response  to treatment  with  probability of response  $p(Y_t=1\mid   A_t=a,  \theta)=\theta_a, a=0, \dots, K.$ 
 We use   independent  beta random  variables  $\theta_a \sim \mbox{Beta}(\theta_a ; v_1, v_2)$ for $a=1, \cdots, K$ to  define  the   prior for $\theta=( \theta_0, \dots, \theta_K).$

{  
	The  trial is designed to generate  estimates of the treatment effects, {$\gamma_a$}, with low  posterior variance  at the end of the trial.  We use an information measure which is consistent with this  goal, 	
	\begin{align}\label{BUD:metric:1}
	u(\Sigma_t)=  \sum_{a=1}^K \Big( v_a - 
	\mbox{Var}( \gamma_a  \mid  \Sigma_t ) \Big),
	\end{align}
	where $v_a=\mbox{Var}(\gamma_a )$. 
}
When the  primary  goal    is  to  test  {  the null hypothesis $\mathcal H_{0,a}: \gamma_a \leq 0$}, a slightly different   utility function 
{$-\mbox{Var}(g(\gamma_a) \mid  \Sigma_t)$}  can  be  considered{, where    $g: [-1,1] \to [0,1]$  is}  a
monotone  function  with  plateaus.   
Each patient is assigned to   treatment $a \in \mathcal A$ with probability
\begin{align}\label{BUD1}
p(A_{t+1} = a\mid  \Sigma _{t} ) 
\propto 
\Bigg( \sum_{a'=1}^K 
\mathbb E \Big[ v_{a'} -  \mbox{Var}(\gamma_{a'}  \mid  \Sigma_{t+1} )  \;\big|\;   A_{t+1}=a, \Sigma_{t} \Big] - 
u( \Sigma _t )   
\Bigg) ^{h(t)}.    
\end{align}

\noindent {\bf Simulation Study:}
We  discuss results of a  simulation study for a 
4-arm trial using the information measure  (\ref{BUD:metric:1})  
to estimate treatment effects.
We considered  four scenarios with constant response rate of   $0.4$ for the control arm (see Table \ref{TAB1}) 
and use uniform prior $ v_1=v_2=1$  for  $\theta_a$, $a=0, \cdots, 3$.  
In  Scenario 1 all   experimental arms are ineffective with response rates equal to 0.4.  
Whereas in the  other  scenarios  some of  the  arms have  positive  treatment  effects.
To simplify comparison to  balanced randomization (BR) we set $T=336$, 
because 
a BR design with one-sided Fisher's exact test  and  type I and II error rates of 0.05 and 0.20 for the alternative 0.6 vs 0.4 requires 84 patients per arm. 
We compare the BUD design to three alternative randomization methods: 
BR,
Bayesian adaptive randomization (BAR)  as described in  \cite{trippa2012bayesian},  and 
the doubly adaptive coin design (DBCD)   \cite{hu2006theory}  targeting   assignment  frequencies  equal to  
the Neyman  allocation
$\rho_a \propto\sqrt{ \theta_a (1-\theta_a) }$ (DBCD1)  or   equal to  $\rho_a \propto \sqrt{ \theta_a}$ (DBCD2)  \cite{hu2006theory}.

Table \ref{TAB1} 
shows  the average  number of patients randomized  to each arm across 5,000 simulations and the 
mean squared error (MSE) of the treatment effect estimates.
We also show the power of Fisher's exact test (without correcting for adaptivity  \cite{Berry1987}) 
for {  the null hypothesis $\mathcal H_{0,a}: \gamma_a \leq 0$}.
Most notable, among all five designs, 
the BUD design has the lowest MSEs  for all experimental arms across all four scenarios.   
When all  experimental arms are ineffective BR,  DBCD1 and DBCD2 randomize on average 84 patients to each arm with standard deviations (SDs) of 8, 4 and 6. 
BAR and the BUD design randomize more patients to the control arm compared to the remaining designs (118, 97).
For BAR this is   expected  by construction of the randomization probabilities  \cite{trippa2012bayesian}.  
In  the BUD design,  
 the assignment of a patient to the control arm reduces 
the  uncertainty on $\theta_0$ and  therefore  leads to    uncertainty reductions
for all  treatment effects $\gamma_a=\theta_a - \theta_0$, $a=1, \dots, K$,
while the  assignment of a patient to  arm $a, a>0, $   reduces   the expected variance only of $\theta_a$ and $\gamma_a$.

When only arm 1 has  a treatment effect (Scenario 2),  
BAR assigns on average 103 and 100 patients to arm 1 and the control arm (SD 10 and 14) and has 87.5\% power.
The BUD, DBCD1 and DBCD1 designs have    82.2\%, 79.8\% and 81.2\%  power  respectively, and 
for  all three designs the standard deviation of the number of  enrolments 
to arm $a=0, \dots,K$  is substantially smaller than for BAR.
  The utility function  (\ref{BUD:metric:1})  targets    treatment effect estimates with low uncertainty. 
One can therefore expect  similar  numbers of patients assigned  to each experimental arm. 
In contrast with BAR  the   arm-specific  sample sizes    tend  to  be markedly different  across experimental treatments.       
In  the    sections \ref{Section:FindBestArm}, \ref{Sec:SubgroupTrial} and \ref{Sec:CoPrimary}   we  discuss  BUD designs  driven by 
 different  information measures  that,  similar  to BAR,  tend  to  increase  the    sample  size  of  the  best
experimental arms. 

{  We also compare the BUD design to an alternative Bayesian design  
\cite{thall2007practical} (BAR2) which assigns patients to  arms with probability proportional to $p( \theta_a > \theta_{a'} \text{ for } a'\neq a | \Sigma_t )^{ t/(2T) }$.  
Compared to the BUD and BAR designs, 
BAR2 assigns   more patients on average to the most effective arm,  
$(84, 161, 180, 130)$ assignments for BAR2 in scenarios 1 to 4 compared to $(80, 103, 117, 83)$ for BAR and 
$(73, 73, 75,  70)$ for the BUD design, respectively.
 BAR2 tends  to have larger MSEs for treatment effect estimates compared to the BUD and BAR designs. 
For  example,  the MSEs of arm 3 in Scenarios 1 to 4 equal 7.3, 9.7, 10 and 9.9 for BAR2 compared to 5.5, 5.4, 4.7 and 5.1 for the BUD design. 
}

\begin{figure}[ht]
\begin{center} 
\includegraphics [width=15cm, height=15cm]  {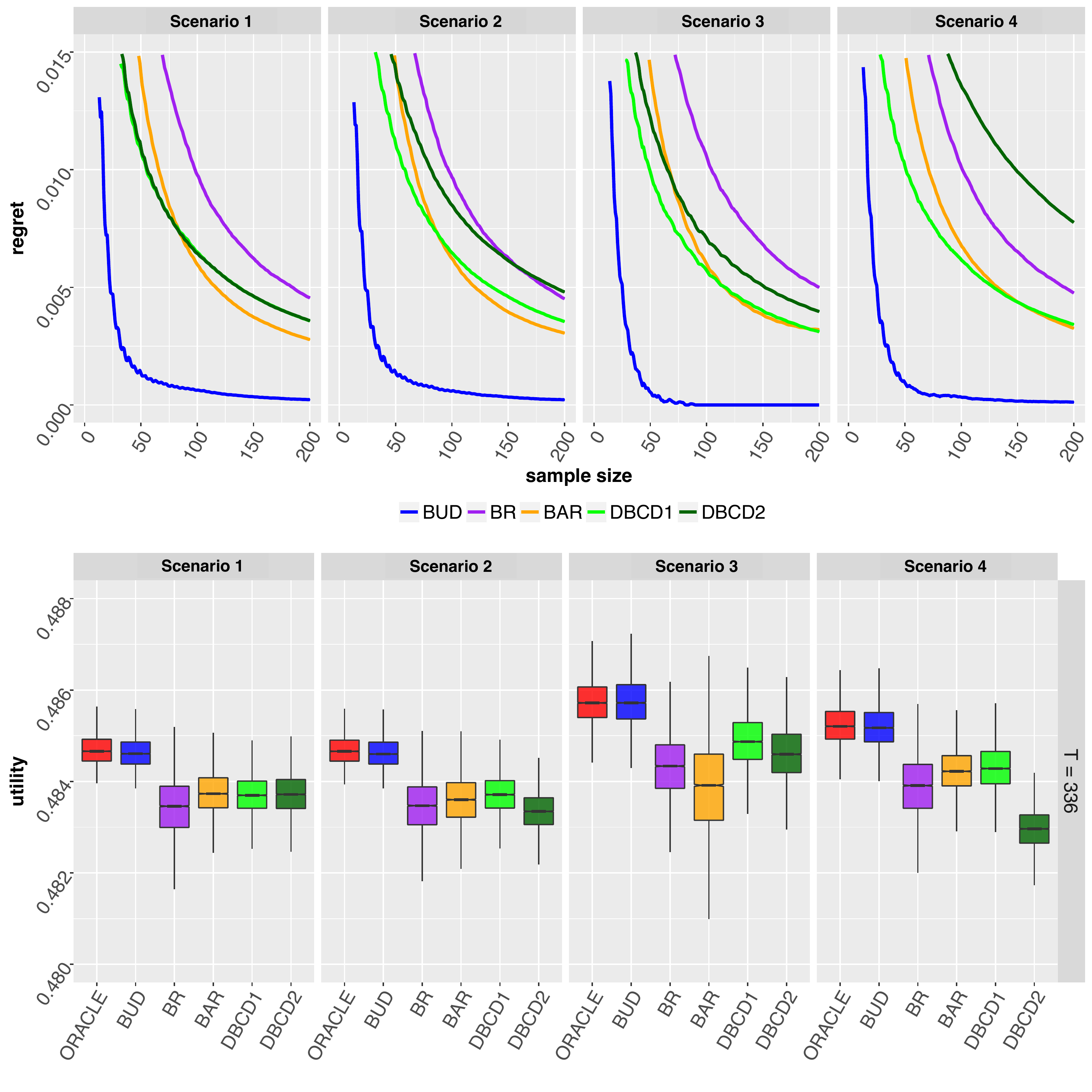}
\caption{ 
Information collected at the end of  a  multi-arm clinical trial with four experimental arms. 
Results are based on 5,000 simulations of a clinical study with four experimental arms using
balanced randomization (BR), Bayesian uncertainty directed (BUD) design,  Bayesian adaptive randomization (BAR) and two versions of the doubly adaptive coin design (DBCD1 and DBCD2). 
The top  panel shown for each of the four  designs the regret function, which is defined as the  
difference between the expected information $\mathbb E[u(\Sigma_T)]$ of the design  and the expected information of the oracle design for different values of $T=1, 2, \dots, 200$.  
The bottom panel shows boxplots of the utility $u(\Sigma_t)=  \sum_{k=1}^K \Big( v_k - \mbox{Var}( \gamma_k \mid  \Sigma_t ) \Big)$, with  prior variance $v_k$, across the simulated trials for a  sample size of $T=336$.
}\label{Fig:utilityMAMS}
\end{center}
\end{figure}

We further  investigated the behavior of BUD design by  repeating  the  simulations over a range of  sample sizes $T$.  
In this  comparison  we  also   included, for each scenario, {     a}   hypothetical   oracle  design. 
For a fixed total sample size $T$, the  oracle design  is   defined by the number of patients that should be assigned to each experimental arm (with sum equal to $T$) in order to maximize $u(\Sigma_T)$  in expectation, assuming  that  the oracle  knows $\theta$.  
We   computed   for each scenario and   $T=1, \dots, 200$,
the combination of sample sizes $\displaystyle \{T_a\}_{a=0}^K,$ $T=\sum_{0 \leq a \leq K} T_a,$ 
that minimizes  the  expected  value  of    $\sum_{a=1}^K  \mbox{Var}(\gamma_a \mid  \Sigma_T) $. 
The top row of Figure \ref{Fig:utilityMAMS} 
shows, for each scenario, the difference between the expected information $\mathbb E_d[u(\Sigma_T)]$ of the four randomization methods and the expected information of the oracle design 
for different values of $T$.  
{  
For the BUD design  we  observe the lowest regret values  across all  sample sizes $T=1, \dots, 200$.  
This is  expected because the BUD design, unlike the other    designs,  is  driven  by  the  utility  $u$,  which is also used to define regret values. 
 We observe that  the regret of the BUD design   is  considerably  closer  to  the  benchmark  utilities of  the  oracle
  design  compared to the  other designs,   and  that the  discrepancy  between the BUD  design and  the benchmark  vanishes quickly as  the  sample  size  increases. 
}%

{ 
 We also explore the sensitivity of BUD designs to different information measures that are consistent with the study aims.
We used the same  scenarios as in the previous paragraphs and consider   BUD designs defined 
by the  entropy, 
	$-\sum_{a=1}^K \mathbb E_{p(\gamma_a|\Sigma_t)}[ \log  p(\gamma_a|\Sigma_t)]$ (BUD-E),  the sum of the mean absolute error of treatment effects estimates   
	 (BUD-MAD), or the  
	 posterior variance and  entropy of the discretized treatment effect $\gamma^c_{a}$ { (BUD-DV and BUD-DE)}, which is defined as $\gamma^c_{a}=0, 1,$ or $2$ when the treatment effect $\gamma_a$ falls into  the  intervals 
	$(-1, 0]$,  $(0,0.25]$ or $[0.25, 1]$. 
	Table \ref{S-TAB:TABSensitivity:utility} in the supplementary material shows the mean squared error (MSE), 
	the average number of enrollments per arm and power for  these  BUD designs. The 
	designs BUD-E and BUD-MAD  have  similar MSEs for the treatment effect estimates across scenarios and assign on average a similar number of patients to each treatment arm.
	{ By contrast}, the truncation of the treatment effect in BUD-DE and BUD-DV increases the variability of treatment effect estimates across scenarios. 
}

{  In  the supplementary material 
we   add to  these  comparisons  the evaluation of alternative BUD designs for controlled multi-arm trials 
where  the  utility  $u$  is representative  of different study aims.
The supplementary material   includes a sensitivity analysis  (Table \ref{S-TAB:DifPrior}) of  the BUD design to variations of the prior model. }

\subsection{Selection of the Best  Experimental Treatment} 
\label{Section:FindBestArm}
Companies and  investigators   often explore multiple experimental treatments  within  a single study to select the  best    treatments   \cite{stallard2008group}.
We consider a K-arm study  without   control arm 
%
to identify treatment   $\displaystyle a^\star = \mathop{\arg \max}_{1 \leq a \leq K} \theta_a$.  

The trial  estimates the response rate $\theta_{a^\star}$  of the most effective therapy $a^\star$ with 
 posterior distribution 
$\displaystyle \theta_{a^\star}$
$$
p_{\theta_{a^\star}}( x  \mid  \Sigma_t) = 
\sum_{a=1}^K \Bigg[  
p_{\theta_a}(x \mid  \Sigma_t) \prod_{j \neq a} p(\theta_j   \le   x \mid    \Sigma_t) \Bigg] ~ \text{ for } x \in (0,1).$$ %
Patients are  randomized  with the aim to minimize  uncertainty on $\theta_{a^\star}$.
Note  that  a  BUD design can   minimize  uncertainty  on   $a^\star$, 
or  it  can  minimize uncertainty on  the unknown  response rate  $\theta_{a^\star}$.

{  	We use a utility function that is consistent with the aim of minimizing uncertainty on $\theta_{a^\star}$ and measure information using the (negative) posterior  entropy of $\theta_{a^\star}$,
	$H[p(\theta_{a^\star}\mid \Sigma_t)] 
	=  \int_0^1  p_{\theta_{a^\star}}(z\mid  \Sigma_t)  \log  p_{\theta_{a^\star}}(z\mid  \Sigma_t )  d z$.
	Other measures, such as the posterior variance of $\theta_{a^\star}$ can  be considered. 
	When the primary goal is to select arm $a^\star$  one may  
	specify  $u$ equal to the entropy of the posterior distribution $p(a^\star  | \Sigma_t)$.
}

Using the randomization scheme of section \ref{GeneralProposal},    
$$
p(A_{t+1} = a\mid \Sigma _{t} ) 
\propto  
\Big( 
  \mathbb E \Big[  H[ p(\theta_{a^\star}\mid \Sigma_{t+1}] \Big|\, \Sigma_{t},  A_{t+1}=a   \Big] - H[ p(\theta_{a^\star}\mid \Sigma_t)]  \Big)^{h(t)}.
$$  

\noindent {\bf Simulation Study:}
We conducted a simulation study for a 4-arm trial that selects  the most effective treatment.
We consider three scenarios (see Table \ref{TAB_best}).
Arm 4 is the most effective treatment in  all scenarios.
We compare the BUD design to BR,  BAR (using Thomson's rule) and the randomized-play-the-winner (RPW) design  \cite{wei1979generalized}.
For all randomization schemes,  at  completion  of  each  simulation
we compute  the probabilities
$\displaystyle p(\theta_{a} =  \max_{a'=1, \dots, 4} \theta_{a'}\mid  \Sigma_T )$ using 
  a uniform  prior  for  
$\theta = (\theta_1, \dots, \theta_4)$ and  select    the  arm $a^\star$  with the highest  posterior probability of 
being the  best  available  treatment.

Table \ref{TAB_best} shows the average number of patients randomized to each  arm across 10,000 simulated trials with $T=30, 50$ or $70$ patients,  and 
provides  additional  operating  characteristics. 
The BUD design selects the most effective arm more frequently  across simulations  compared to the  three alternative designs.
In scenario 1 after $T=30$ assignments, for  example, in 57\% of the simulated trials 
the BUD design selects    the right arm, compared to 52\%, 56\% and 53\% for BR, BAR and RPW, respectively. 
The BUD  design assigns on average the same number of patients to each arm as BAR, but has lower assignment variability.
BAR is also similar  to the BUD  design when we consider   the  probabilities  of  selecting  the  best  available  treatment,
but it has the largest assignment variability among all randomization schemes. 
As shown in the last column of Table \ref{TAB_best}, 
across all scenarios  the BUD design has the lowest mean squared 
error  for the estimate of  $\max_a \theta_a$.

We also derived  optimal sequential Bayesian designs $d^\star$ in (\ref{d:OB}) using BI for comparisons.
The  entropy $H(\theta_{a^\star}\mid \Sigma_T)$  defines the  utility function and the  expected utility $\mathbb E_{d^\star}[u(\Sigma_T)]$.
We considered  small sample sizes $T$, up to 30 patients,  for which BI is  feasible.   
These  computations  were followed by  a comparison based  on   $10^5$ simulated  trials  for  BUD, BR BAR, and the RPW designs.    
In each simulation, patients respond to treatments with probabilities  $\theta=(\theta_1, \dots, \theta_4) \sim p(\theta)$  from the
   prior (independent beta)  that  we used to compute $d^\star$. 
We  estimate the expected utility of each design $d$
and the regret $ \mathbb E_{ d^\star}[u(\Sigma_T)]-\mathbb E_{d}[u(\Sigma_T)]$, see Figure \ref{S-Bandit}.
The regret of the BUD design remains close to zero across all sample sizes $T$ that we considered, 
whereas the regret of BR and RPW designs are  larger; for instance  
the  regret  for  BUD, BAR, BR and RPW designs
  is  equal to $ (0.014, 0.035, 0.15, 0.12)$ and 
$(0.017,0.047, 0.30, 0.22)$, with $T=10$ and  $30$ patients, respectively.

{  
We  evaluate the sensitivity of  BUD designs 
  to the choice  of the randomization parameter $h$. 
We conducted  simulations under the  three scenarios  in Table \ref{TAB_best} using 
either $h(t)=1, 2, 6, 12$ or 20.     
The  parameter $h$ in our simulations has minor influence on  the resulting operating characteristics, see supplementary Table \ref{S-TAB:Sensitivity:h}.   
Across  scenarios  the average number of treatment assignments to each arm and 
the proportion of simulated trials that recommend the best arm for further investigation  remains nearly identical when we vary $h$.
}

\section{Biomarker-Stratified  Clinical Trials}\label{Sec:SubgroupTrial}

The development of anti-cancer treatments  focuses increasingly on therapies that target specific  genetic alterations.
Consequently, there has been an increasing interest in biomarker-driven studies  
  \cite{Lee2010, Xu2016, Ventz2017, trippa2017}.
These studies enroll patients with multiple genomic abnormalities in a single multi-arm trial,  
and identify subgroups of patients that respond  to experimental  treatments. 

We consider a  design that evaluates $K$ experimental therapies  in biomarker subgroups.   For each patient $t$ the   
vector $X_t \in \{0,1\}^B$ defines the patients' biomarker  profile, here  $X_{t,\ell}= 1$  if  patient $t$ has a positive marker status for biomarker  $\ell=1, \dots, B$ and zero otherwise.
In our examples $B$ will be equal to 2,3 or 4.
Patients  can  be  partitioned  into $2^B$ groups.
We assume binary endpoints with  probability of response 
$p(Y_t=1\mid X_t=x, A_t=a, \theta) =\theta_{x,a}$  
  for patients with profile $x \in  \{0, 1\}^B$.  
Here $\theta_{x,0}$ is   the subgroup-specific response probability  for the control  arm $a=0$.
%
%
%
%
%

In oncology, 
treatments are commonly developed for patients with specific genomic abnormalities 
 \cite{polivka2014molecular}.
Pre-clinical studies typically suggest effects of a treatment $a$ in a targeted biomarker subgroup.
But in  many cases positive effects  in other groups of patients 
 are hypothesized, and treatments are evaluated in a population larger that the target groups  \cite{polivka2014molecular}.     
%
%
We assume that  treatment $a>1$ targets biomarker $1 \leq b_{a} \leq B$, 
and it  is therefore {\it a priori} more likely to be effective for patients with positive biomarker $X_{t,b_a}=1$ status than  for other subgroups.

%
%
%
%
%

%

The  indicator 
$ \displaystyle E_{\ell,a} =  \mathbf{1} \Big (\bigcup_{x \in \{0,1\}^B: x_{b_a} = \ell  } \{  \theta_{x,a} >\theta_{x,0} \}\Big)$
corresponds, for $\ell=1$ and $0$, to  treatment effects for  biomarker $b_a$ positive  and negative patients, respectively. 
%
We first specify the prior  probabilty $
p(E_{1,a}=1) 
=\pi$.
%
Since treatment effects for patients   without genomic alteration $b_a$ can not be ruled out 
we specify
$p(E_{0,a} =1\mid  E_{1,a}=1) 
=\lambda$.
Lastly, since drug $a$ was developed primarily for biomarker $b_a$  we assume 
$ p(E_{0,a} =1 \mid  E_{1,a}=0)
=0$.

%
%
%
%
%
%
%




The prior 
distribution for 
$\theta = (\theta_{x,a}; x \in \{0,1\}^B, 0\leq a \leq K)$
conditional on the indicators $E_{\ell,a}$  is a product of Beta$(1,1)$ distributions restricted to the subset $R\subset [0,1]^B$ consistent with these indicators, 
\begin{equation}\label{model}
p \Big( \theta  \, \Big | \,     E_{\ell,a}; \ell=0,1, a=0, \dots, K   \Big) 
\propto    I( \theta \in R ) 
  \prod_{ \substack{ x \in \{0,1\}^B, \\ a=0, \dots,   K }} \text{Beta}(\theta_{x,a}\,;\, 1,1).
\end{equation}
Alternative  prior distributions, different   from the simple Bayesian model (\ref{model}) that we specified, could   be   applied  
within the BUD framework, see for instance  \cite{Xu2016}.

We define a BUD design 
 that evaluates  treatments  in the targeted  subgroups, and  explores   possible additional treatment effects  in the biomarker negative subgroups. 
The information measure is   
\begin{align}\label{BUD:Biomarker}
u\!\big(\Sigma_t\big)
& = \! -\sum_{a=1}^K \Bigg \{ 
H_\text{as} \big[ p( \{E_{1,a}=1 \} \! \cup \! \{  E_{0,a}=1\}\mid  \Sigma_t) \big] \! +  \nonumber \\
& ~~~~~~~~~~~~~~~~~~~ w \times\bigg ( H_\text{as}  \big[ p(E_{1,a}=1\mid  \Sigma_t)  \big]+H_\text{as} \big[ p(E_{0,a}=1\mid  \Sigma_t) \big] \bigg )
\Bigg \}.
\end{align}
Here  $H_\text{as}[p ] = p - p^\beta$, with $ \beta>1$, is the asymmetric  entropy 
 \cite{lenca2010construction} and
{ 
  $w>0$ weights the information gains in the  subgroups   and  in  the overall population. Allocations in  the BUD design using (\ref{BUD:Biomarker}) are driven by the expected reduction of the  posterior entropy of positive treatment effects. 
The parameter $\beta$ controls the curvature {      $H''_\text{as}[p] \propto -p^{\beta-2}$ } of the entropy, 
and, {      using a Taylor approximation,  the expected  variation  in  the entropy 
of $p(E_{\ell,a}=1 | \Sigma_t)$ } given  $A_{t+1}=a$  equals approximately 
$$H''_\text{as}( p(E_{\ell,a}=1| \Sigma_t) ) \times \mbox{Var}[ p(E_{\ell,a}=1 | \Sigma_{t+1}) | \Sigma_t, A_{t+1}=a ]/2.$$  
Values of $\beta>2$  tend to  reinforce  the randomization probabilities of the  arms that  are  more  promising 
 accordingly to  the  posterior probabilities  $ p(E_{\ell,a}=1| \Sigma_t)$ and $p( \{E_{1,a}=1 \} \! \cup \! \{  E_{0,a}=1\}\mid  \Sigma_t)$.    
}

{     In contrast to previous BUD designs in Section \ref{Exa1}, which focus on  treatment effects estimation, the  metric   (\ref{BUD:Biomarker})  quantifies  uncertainty on the  presence or absence of  treatment effects in  biomarker targeted subgroups and in the overall population. This is consistent with the study aim of testing  efficacy. 
}

The BUD design assigns patients to  treatment  $a$ with  probability   
$$
p \big( A_{t+1} = a\mid    X_{t+1}=x, \Sigma _t \big) 
\propto
\Bigg[ \mathbb E \Big[ u (\Sigma_{t+1}) \;\Big | \; X_{t+1}=x, A_t=a, \Sigma_t \Big] - u \big( \Sigma_t \big)\Bigg]^{h(t)}, \text{ for } a=0, \dots, K. 
$$

\noindent {\bf Simulation Study:}
Table  \ref{TAB::Biomarker} 
 summarizes the results of a simulation study  for a controlled biomarker trial with four experimental treatments and 
an overall sample size of $T=500$ patients. 

We considered five scenarios with four (Scenarios 1-3), three (Scenario 4) or two (Scenario 5) biomarkers $B$.  
The true response probability for the control equals $\theta_{x,0}=0.35$ across all five scenarios and subgroups $x\in \{0,1\}^B.$
Treatments $a=1, \dots, 4$ target biomarker $b_a=a$ in the first three scenarios.  
In  Scenario  1 there are no  positive treatment effects (PTEs), and for all  therapies $a>0$ the  probabilities $\theta_{x,a}, x\in\{0,1\}^B$ are identical to the control. 
Whereas in Scenarios 2-3, treatment $a=1$, and treatment $a=2$ (only in Scenario 3), have  PTEs with  $\theta_{x,a}=0.55$  for all patients with positive  biomarker  
status  for $b_a=a$, i.e. $X_{t,b_a}=1$.
%
In Scenario 4, treatments $a=1, 2$ both target biomarker $b_a=1$ and have a PTE on this subgroup  with 
$(\theta_{x,1}, \theta_{x,2}) = (0.55, 0.65)$. The remaining  treatments $a=3,4$,  which target biomarker 2 and 3, have no PTEs.
Lastly, in Scenario 5, therapies $a=2,3,4$ have PTEs in their target population $b_a =1,2,2$ with $\theta_{x,a}=0.55, 0.55$ and $0.65$ for all  biomarker 
positive patients $X_{t,b_a}=1$. Additionally,  the PTE  of  treatment  $a=1$ ($b_1=1$) extends  to  all patients irrespectively of the  profile $X_t = x$, with   $\theta_{x,1}=0.5$. %

We used $(w,\beta)=(5,6)$ to define the utility function of the BUD design, 
and compared the design to  a balanced design (BR) that randomizes patients to the five  arms  with equal probabilities.
For each treatment $a>0$, we tested treatment effects in 
the biomarker-targeted  group  (patients with $X_{i,b_a}=1$)  and  in the  biomarker 
negative group ($X_{i,b_a}=0$)   using a bootstrap test similar to \cite{trippa2012bayesian, Cellamare2016, Ventz2017b}, accounting for adaptive randomization.

Table  \ref{TAB::Biomarker}  shows  mean and  variability for  the  allocation of patients to treatments, and the power of detecting  treatment effects. 
  The targeted type I error rate is set at $10\%$.   
Similar to the multi-arm BUD designs  in section \ref{Exa1}, the biomarker  BUD design allocates substantially more patients to the control arm than the BR design  to  reduce  uncertainty  
on treatment effects.
This translates into a higher power across all  scenarios compared to the BR design.       
In  Scenario  1, without  treatment effects, the BUD design assigns on average a higher number of patients with targeted biomarker profile $X_{b_a}=1$ to  the corresponding experimental treatment  $a=1, \dots, 4$ than the BR design.  It tends to match   targeted treatments   and  biomarker  profiles. 
This is due to the specification of a higher a priori probability of  PTEs for the targeted biomarker-subgroups. 
Interestingly, 
in  Scenario  4, where  arms $a=1$ and $2$ have PTEs in the first biomarker subgroup, 
the BUD design on  average assigns more patients to the arm with the smaller treatment effect, which  is more difficult to identify. 
This translates into a gain in power of 11\% compared to BR.
Similarly, in  Scenario  5, 
where all treatments $a>1$ have PTEs
the BUD design randomizes on average more patients  to experimental arms  with small treatment effects.    
This translates 
into substantial  gains in   power   compared to  BR. 
   
In contrast to BAR, under a BUD design with utility (\ref{BUD:Biomarker}),
the average arm-specific sample  size is  not expected  to  increase 
with the  magnitude  of the  treatment effect. 
Indeed with large treatment effects, 
 uncertainty on  the  indicators  $E_{1,a}$ and $E_{0,a}$  in the targeted and non-targeted subgroups
  vanishes rapidly  with  a  limited  number of observations.
  
{  
We evaluated the sensitivity of the BUD design  to different choices of the parameters $(\beta,w)$ which  define the uncertainty function (\ref{BUD:Biomarker}). 
We repeated the simulations under scenarios 2 and 4 using either $w=5$ with $\beta = 2,6,8$ 
or  using $w=2,5,8$ with $\beta=6$ (Tables \ref{S-TABSensitivityBiom} and \ref{S-TABSensitivityBiom2}).
A BUD design with symmetric information measure $\beta=2$   
assigns on average more patients to ineffective experimental arms than using $\beta=6$ or $\beta=8$. 
The latter two choices have  similar operating characteristic; compared to $\beta=2$, a BUD design with $\beta \geq 6$ assigns on average  fewer patients to ineffective agents and increases the power by 11\% for arm 1 in scenario 2, and {     by} 5\% to 9\% for arms 1 and 2 in scenario 4  
(scenario 2: 74\%, 86\% and 85\% power for $\beta=2,6,8$  using  Fisher  exact test).   
As shown in Tables  \ref{S-TABSensitivityBiom} and \ref{S-TABSensitivityBiom2} the BUD design is relatively insensitive to the choice of  $w$.
}

\section{Trials with Co-Primary Endpoints}\label{Sec:CoPrimary}
In several   contexts,  such as  Alzheimer's  disease, 
a single  endpoint has been shown to be insufficient to capture  patients' 
response to treatments adequately  \cite{druss2001integrated}.     
Several  authors  recommend  to evaluate   new  treatments using  
multiple clinical endpoints  \cite{druss2001integrated}.     
Trial designs that  evaluate   treatments  using  multiple outcomes have been proposed in  \cite{teixeira2009statistical, thall1995bayesian}.  
We consider a BUD design 
for a controlled multi-arm trial  that evaluates $K$ experimental treatments 
using  two binary endpoints $Y_t = (Y_{t,1}, Y_{t,2})$,
for example, in  Alzheimer's  disease,   the  
cognitive performance  and
the  physical status  after  treatment. 
Regulatory guidelines  \cite{druss2001integrated} 
recommend  the investigator  to demonstrate superiority of   the experimental treatment $a>0$ 
compared to the control arm  $a=0$ on both endpoints.

For each treatment $a$,   
$p(Y_t = y \mid A_t=a) = \theta_{y, a}\ge 0$, with $y\in \{0,1\}^2,  $ and  $\sum_{y \in \{0,1\}^2}\theta_{y, a}=1.$ 
We  use independent Dirichlet prior  distributions for  the  arm-specific parameters 
 $\theta_a = \big( \theta_{(1,1),a}, \allowbreak \theta_{(1,0),a}, \allowbreak \theta_{(0,1),a}, \allowbreak  \theta_{(0,0),a}  \big)$.  
The  parameters  $\theta_a$   specify       
 the marginal probabilities  $(\nu_{1,a}, 
\nu_{2,a})$  of  the two endpoints.
We  use 
$\gamma_{\ell,a} =  \nu_{\ell,a}- \nu_{\ell,0},\;\; \ell =1,2,$ to  indicate treatment effects for both endpoints, and  define  the indicators
$E_{\ell,a}     = \mathbf 1 \{ \gamma_{\ell,a} >0 \},$  $\ell=1,2$, 
and  
$E_{a} = E_{1,a} \times E_{2,a}$.

The posterior probability $p(E_{a}=1 \mid \Sigma_t)$ summarizes the  available  evidence of therapy $a$ as suitable alternative to the control therapy.  
{  
		Although  the  investigator's focus  is on   testing   $E_{a}=0$  versus   $E_{a}=1$,  improvements for one of the two outcomes  are  also  relevant. 
	We therefore consider a composite uncertainty measure 
	\begin{equation}\label{utility_1_multi}
	u_1(\Sigma_t) = -\sum_{a=1}^K \Bigg\{
	H_\text{as}[p( E_a=1 | \Sigma_t) ] +
	w 	\times 
	\bigg( H_\text{as}[p( E_{1,a}=1 | \Sigma_t)] +
	H_\text{as}[p( E_{2,a} =1| \Sigma_t)] \bigg) \Bigg\},
	\end{equation}
	where as before  $H_\text{as}[ p ] = p - p^\beta$, $\beta>1$ and $w>0$, which  is consistent with the study goals of testing  $E_a$ and 
	 improvements  for single endpoints $E_{a,\ell}, \ell=1,2$. 
}

To explore variations  of the BUD design's operating characteristics with different information measures we also considered an alternative utility function
\begin{equation}\label{utility_2_multi}
u(\Sigma_t) = -\sum_{a=1}^K \Bigg\{
 \mbox{Var}(\gamma_{a}| \Sigma_t ) + 
w \times\Big( \mbox{Var}(\gamma_{1,a}| \Sigma_t ) +
\mbox{Var}(\gamma_{2,a}| \Sigma_t ) \Big ) \Bigg\},
\end{equation}
where $\gamma_{a}$ is the  difference between arm $a>0$ and the control arm in the probability of an individual's positive response on both endpoints.

\noindent {\bf Simulation Study:} We  discuss  a simulation study for a  multi-arm trial with  
four experimental arms and an overall sample size of $T=348$ patients.
For  the arms $a=0,2,3,4$ the  parameters  
$\theta_a = (0.15, 0.25, 0.4, 0.2)$   are  identical  across  the  four scenarios that we considered. 
In  Scenario 1, 
 arm $a=1$ has no PTE    while   in  Scenarios 2--4  $(\gamma_{1,1}, \gamma_{2,1})=(0.2,0.2), (0.2,0.05)$, and $(0.05,0.2)$ 
respectively.  
%

Table \ref{S-TABmulti} in the supplementary material, 
illustrates  the  MSE of the  effects estimates 
$\widehat \gamma_{\ell, a} $, 
 power for  null hypotheses 
$\mathcal H^{\ell}_{0,a}:  \gamma_{\ell, a}\leq 0,$  $\ell=1,2$  and  the proportion of
 simulations in which both hypotheses are rejected 
for  BUD designs with utility function (\ref{utility_1_multi}) ($\mbox{BUD}_1$)  or  utility function
(\ref{utility_2_multi}) ($\mbox{BUD}_2$).
We also  compared  BUD designs to a balanced randomized (BR) design  that assigns patients to treatments with equal probabilities.  
As expected,  the $\mbox{BUD}_2$ design, which seeks to minimize the posterior variance of the treatment effects,  attains   the lowest  MSEs.
 The BUD  designs  have  nearly  identical power and 
 BR has across all alternative scenarios the lowest power.  
 
 {  
We also  computed  the Bayesian optimal design $d^\star$  using BI
with utility defined by (\ref{utility_1_multi}) for a trial with $K=2$ experimental arms. 
We compare the expected utility of the optimal design  $d^\star$ 
to the expected  utility of the BUD and BR designs  $d=d_{BUD}, d_{BR}$.
 Expected  utilities  are computed  by  integrating  with respect  to the  prior.  
With  co-primary  endpoints 
 dynamic programming becomes  quickly  
computationally challenging
for moderate sample sizes $T \approx 25$, and realistic sample sizes 
 $T\approx 378$ are infeasible.
For values of $T=5, 10, 15$, 
we observe a reduction  of  expected  utility  equal  to (0.59\%, 1.21\%, 2.04\%)
for  BUD  and   (36.92\%, 61.39\%, 77.4\%) for BR.
 }

\section{Discussion}\label{Discussion}
{  There are competing objectives in the design of clinical trials.
These  include  identification of  treatment effects, 
accurate estimates,
and  limiting the number of patients exposed to suboptimal treatments. 
Bayesian adaptive designs { usually }   target the latter aim, or  attempt to balance  competing aims, by unbalancing randomization toward more promising arms.
 In contrast, BUD designs target an information measure $u$ consistent with the goal of testing or estimating treatment effects.
}

{ 
Bayesian adaptive randomization (BAR) has been applied in various clinical studies and, in several settings,  it  has  better operating characteristics  than  alternative designs  \cite{Berry1995, thall2007practical, Lee2010, trippa2012bayesian, Cellamare2016, Ventz2017}. 
   Limitations  of  BAR  have  also been discussed in the literature. 
It is known, for  example, that in  two-arm trials  BAR may increase the overall sample size compared to balanced randomization (Korn and Freindlin, 2011). 
  In multi-arm trials  BAR can be highly variable in the  arm-specific enrollment proportions
   and in some cases  assigns  
with  non negligible probability
more patients to inferior arms than  balanced randomization  \cite{thall2007practical, thall2015statistical}. 
This suggests that investigating alternative approaches to Bayesian adaptive trials, such as BUD designs, remains a relevant problem.

Both, BAR and BUD designs translate  posterior distributions into randomization probabilities. 
These  two  approaches can also be described as  randomized myopic decision rules.  
BAR approximately  optimizes the number of positive  outcomes   \cite{Berry1995}, 
while   BUD designs approximately optimizes  posterior summaries, such as  the variance or entropy, that quantify uncertainty on key  parameters.  
In the two-arm case, \cite{Berry1995} showed that BAR has, in several  scenarios,  operating characteristics similar to the Bayesian  optimal design   when the utility  function coincides with the number of positive  outcomes.
Unlike BAR, BUD designs focus on  information measures  based on the primary  aims of the trial,  such as testing  efficacy  or  estimating  treatment effects.
}

We propose several information measures which reflect the aims of multi-arm trials with biomarker
subpopulations and multiple endpoints, and show that in a number of simulation scenarios, 
frequentist operating characteristics are improved with respect to other strategies. 
In fact, the utility at the end of the experiment is often near the optimal value produced by intractable backward induction algorithms.
The simplicity of the randomization probabilities makes it possible to derive asymptotic limits in certain examples.

The  computing time to simulate a BUD trial increases linearly 
with respect to the sample size $T$ and with respect to the number of  possible actions.
In our examples the computing time for  BUD simulations, as expected, is very similar to BAR, 
and generally  significantly faster  than  computing  the optimal Bayesian design  obtained  with   dynamic programming.

{  
{ A crucial { aspect} of BUD designs is the selection of  information metrics $u$ that represent the primary aims of the clinical study. }
We can classify studies into  broad categories based on their primary aims;
(i)  estimating   treatment effects, 
(ii) identifying  therapies with positive effects, and    
(iii) studies with multiple objectives, 
for instance identifying relevant treatment effects on multiple endpoints or within  subgroups.

 { In case (i),}  the focus  is  on key parameters of interest
and a symmetric uncertainty  metric, such as the entropy or posterior variance, 
can be  used to represent estimation accuracy. 
The sensitivity analyses  in Section 3.2 shows that BUD  designs with these  information metrics have  similar  operating characteristics.

 { In case (ii),} as in  hypothesis testing,  the  parameter  space  is partitioned  into  regions  with  and  without  treatment effects.  
We can use discrete random variables $E_a \in \{0,1\}$ 
or  monotone  functions $g(\gamma_a)$  representative  of  this partition.
 The posterior variance or (asymmetric) entropy    of these random variables 
 can be used to define  uncertainty  measures. 
For example, in a multi-arm study with two experimental drugs and binary response to treatment, 
$\gamma_a = \theta_a - \theta_0, a=1,2$ indicate the difference between the response rate of experimental arm $a$ and the control.
The entropy  of   $E_a = I(\gamma_a>0)$ or variance of  $g(\gamma_a)= \gamma_a / (1+\gamma_a^2)^{1/2} $ can be adopted as uncertainty metrics.

{ In case (iii),}   when the study has  multiple  related goals, a  composite  information measure $u = \sum_j w_j u_j$ 
can  be  used  to  weight different criteria $u_j$.
{  For  example in Section 5, the BUD  design for a trial with two endpoints  is  based  on a  composite utility function. }
}

\bibliography{Bibliography}

\clearpage

\clearpage

\clearpage 
\begin{table}[ht]
\centering
\small

 \renewcommand{\arraystretch}{1.1}
\begin{tabular}{| r  r  |  rrr |  rrr | rrr |}
\hline

 & \textcolor{black}{Control} &  \multicolumn{3}{c|}{Arm 1} & \multicolumn{3}{c|}{Arm 2} & \multicolumn{3}{c|}{Arm 3} \\
 
 \textbf{Design}    & ESS (SD) 
 & \multicolumn{3}{c|}{ {\footnotesize  ESS (SD)  Power  MSE}}  
 & \multicolumn{3}{c|}{ {\footnotesize ESS (SD)  Power  MSE}}  
 & \multicolumn{3}{c|}{ {\footnotesize ESS (SD)  Power  MSE}}    \\
 \hline
&  \multicolumn{10}{l|}{\textbf{Scenario 1}: no effective arm 
$(\theta_0, \theta_1,\theta_2,\theta_3)=(0.4 ,0.4,0.4,0.4)$}\\
  BR        &  84 (8)   & 84 (08) & 04.0 & 5.89 & 84 (8) & 04.1 & 5.86 & 84 (8) & 03.2 & 5.74 \\ 
  BUD     & 118 (3)  & 73 (03) & 03.8 & {\bf 5.44} & 73 (3) & 03.8 & {\bf5.43} & 73 (3) & 04.0 & {\bf5.52} \\ 
  BAR     &  97 (8)   & 80 (21) & 03.9 & 7.33 & 79 (21) & 03.6 & 7.30 & 80 (21) & 03.8 & 7.28 \\ 
  BAR2  & 84  (23) & 84 (22) & 03.9 & 7.26 & 84 (22) & 04.0 & 7.17 & 84 (23) & 04.2 & 7.37 \\
   
  DBCD1 & 84 (4) & 84 (4) & 03.8 & 5.93 & 84 (4) & 03.8 & 5.85 & 84 (4) & 03.6 & 5.82 \\ 
  DBCD2 & 84 (6) & 84 (6) & 03.6 & 5.89 & 84 (6) & 03.9 & 5.92 & 84 (6) & 03.5 & 5.82 \\ 
 \hline
  
& \multicolumn{10}{l|}{\textbf{Scenario 2:}  one superior arm $(\theta_0, \theta_1,\theta_2,\theta_3)=(\textbf{0.4} ,\textbf{0.6},0.4,0.4)$}\\
  BR        & 85 (8) & 84 (8) & 78.6 & 5.75 & 83 (8) & 04.2 & 5.75 & 84 (8) & 03.3 & 5.80 \\ 
  BUD     & 118 (3) & 73 (3) & 82.2 & 5.48 & 73 (3) & 03.6 & {\bf5.46} & 73 (3) & 03.8 & {\bf5.40} \\ 
  BAR     & 100 (10) & 103 (14) & {\bf 87.5} & {\bf 4.89} & 67 (18) & 03.2 & 7.65 & 66 (18) & 03.6 & 7.96 \\ 
  BAR2  & 58  (17 ) & 161 (28) & 85.1 & 6.80 & 58  (17) & 3.4 &  9.80 & 58  (17)  & 3.5  & 9.65\\
  DBCD1 & 84 (4)    & 84 (4)     & 79.8 & 5.99 & 84 (4) & 03.9 & 6.07 & 84 (4) & 04.0 & 6.02 \\ 
  DBCD2 & 80 (6)    & 97 (6)     & 81.2 & 5.63 & 80 (6) & 03.4 & 6.16 & 80 (6) & 03.7 & 6.09 \\ 
    \hline
    
& \multicolumn{10}{l|}{\textbf{Scenario 3:} one superior and one inferior arm  $(\theta_0, \theta_1,\theta_2,\theta_3)=(\textbf{0.4} ,\textbf{0.6},0.4,0.2)$}\\
 BR & 84 (8) & 84 (8) & 78.9 & 5.92 & 84 (8) & 03.6 & 5.70 & 84 (8) & 0.00 & 4.72 \\ 
  BUD & 122 (4) & 75 (3) & 84.3 & 5.10 & 75 (3) & 03.6 & {\bf5.22} & 63 (5) & 0.00 & {\bf4.68} \\ 
  BAR & 111 (10) & 117 (13) & {\bf 90.6} & {\bf4.40} & 75 (20) & 04.0 & 7.22 & 32 (11) & 0.00 & 8.12 \\ 
  BAR2 & 62(18) & 180 (27)& 88.2  & 6.24 & 62 (18)  & 3.4  & 9.04  & 31 (8)  0.0 & 9.95 \\
   
  DBCD1 & 88 (4) & 88 (4) & 81.1 & 5.69 & 88 (4) & 03.7 & 5.66 & 72 (7) & 0.00 & 5.20 \\ 
  DBCD2 & 85 (7) & 104 (6) & 83.8 & 5.25 & 85 (7) & 03.8 & 5.95 & 61 (8) & 0.00 & 5.76 \\ 
  \hline
& \multicolumn{10}{l|}{\textbf{Scenario 4:} three superior arms  $(\theta_0, \theta_1,\theta_2,\theta_3)=(\textbf{\textcolor{black}{0.4}},\textbf{0.6},\textbf{0.65},\textbf{0.7})$}\\
 BR & 84 (8) & 85 (8) & 79.4 & 5.89 & 84 (8) &   92.5 & 5.86 & 84 (8)   & 98.6 & 5.74 \\ 
  BUD & 120 (3) & 74 (3) & {\bf 83.6} & {\bf5.34} & 72 (3) & {\bf 95.1} & {\bf5.18} & 70 (4)  & {\bf 99.1} & {\bf5.14} \\ 
  BAR & 90 (4) & 80 (9) & 79.2 & 7.33 & 82 (8) & 93.4 & 7.30 & 83 (8)   & 98.6 & 7.28 \\ 
  BAR2 & 35 (10) &  75  (22) & 55.1 & 11.94 & 96  (26) & 78.8 & 10.77 & 130  (29) & 95.3 & 9.87\\
  DBCD1 & 86 (4) & 86 (4) & 80.4 & 5.93 & 84 (5) & 94.0 & 5.85 & 80 (5) & 98.6 & 5.82 \\ 
  DBCD2 & 70 (6) & 85 (5) & 75.8 & 5.89 & 89 (5) & 91.7 & 5.92 & 92 (5) & 98.4 & 5.82 \\       
  \hline

\end{tabular}

\caption{\small Expected sample size (ESS), standard deviation (SD), power and mean squared error (MSE), for each experimental arm in a 4-arm trial with $336$ patients using 
either
balanced randomization (BR),
Bayesian uncertainty directed (BUD) design,
{Bayesian adaptive randomization (BAR) as described in    \cite{trippa2012bayesian}
and the version described in \cite{thall2007practical} (BAR2),}
and two versions of the  doubly adaptive biased coin design, DBCB1 and DBCD2.
The MSE is scaled by a factor of $10^3$.
}  
\label{TAB1}
\end{table}

\clearpage 

\clearpage
\begin{table}[ht]
\centering
\footnotesize

 \renewcommand{\arraystretch}{1.1}
\begin{tabular}{| ll | rr | rr | rr | rr | r |}
\hline

 & & \multicolumn{2}{|c}{Arm 1}  &  \multicolumn{2}{|c|}{Arm 2} & \multicolumn{2}{|c|}{Arm 3} & \multicolumn{2}{c|}{Arm 4} &\\
 \textbf{T}  & \textbf{Design}    & \multicolumn{2}{|c|}{ESS (SD)  $P_1$  }  & \multicolumn{2}{|c|}{ESS (SD)  $P_2$  }   &  \multicolumn{2}{|c|}{ESS (SD)  $P_3$  }   
 & \multicolumn{2}{|c|}{ESS (SD)  $P_4$  }   & MSE\\
 \hline
   \hline
 \multicolumn{4}{|c}{\textbf{Scenario 1}} & \multicolumn{7}{l|}{$(\theta_1,\theta_2,\theta_3,\theta_4)=(0.3,0.4,0.5,\textbf{0.6})$}\\
30 & BR & 7 (3) & 0.059 & 8 (3) & 0.155 & 8 (2) & 0.269 & 8 (4) & 0.517 & 13.63 \\ 
   & BUD & 4 (10) & 0.035 & 6 (8) & 0.117 & 8 (10) & 0.274 & 12 (12) & \textbf{0.573} & \textbf{8.83} \\ 
   & BAR & 4 (12) & 0.044 & 6 (7) & 0.123 & 8 (7) & 0.273 & 13 (13) & 0.560 & 9.16 \\ 
   & RPW & 7 (4) & 0.056& 7 (3) & 0.133 & 8 (3) & 0.280 & 8 (4) & 0.531 & 11.9 \\
     \cline{2-11}    
50 & BR & 12 (3) & 0.033 & 13 (3) & 0.101 & 12 (3) & 0.276 & 13 (5) & 0.590 & 9.22 \\ 
   & BUD & 6 (15) & 0.018 & 9 (12) & 0.075 & 13 (16) & 0.247 & 22 (15) & \textbf{0.659} & \textbf{6.35} \\ 
   & BAR & 5 (16) & 0.024 & 8 (15) & 0.086 & 13 (10) & 0.251 & 24 (21) & 0.640 & 6.80 \\ 
   & RPW & 11 (4) & 0.031 & 12 (4) & 0.091 & 13 (3) & 0.2592 & 14 (6) & 0.619 & 8.23 \\ 
      \cline{2-11}
 70 & BR & 17 (3) & 0.015 & 18 (4) & 0.079 & 17 (4) & 0.263 & 18 (5) & 0.643 & 7.65 \\ 
   & BUD & 7 (23) & 0.008 & 11 (18) & 0.054 & 18 (20) & 0.222 & 34 (15) & \textbf{0.715} & \textbf{5.10} \\ 
   & BAR & 6 (23) & 0.011 & 10 (25) & 0.060 & 18 (16) & 0.226 & 36 (30) & 0.703 & 5.31 \\ 
   & RPW & 15 (6) & 0.013 & 17 (5) & 0.067 & 18 (4) & 0.240 & 20 (9) & 0.680 & 6.77 \\ 

    \hline

 \multicolumn{4}{|c}{\textbf{Scenario 2}} & \multicolumn{7}{l|}{$(\theta_1,\theta_2,\theta_3,\theta_4)=(0.4,0.4,0.4,\textbf{0.8})$}\\

 30 & BR & 7 (3) & 0.049 & 8 (3) & 0.058 & 8 (2) & 0.040 & 8 (4) & 0.853 & 13.95 \\ 
   & BUD & 3 (10) & 0.024 & 4 (11) & 0.026 & 3 (13) & 0.028 & 20 (12) & \textbf{0.921} & \textbf{11.98} \\ 
   & BAR & 3 (12) & 0.035 & 3 (8) & 0.036 & 3 (13) & 0.033 & 20 (13) & 0.894 & 13.79 \\ 
   & RPW & 7 (4) & 0.041 & 7 (2) & 0.039 & 7 (7) & 0.041 & 10 (2) & 0.878 & 13.47 \\ 
     \cline{2-11}
 50 & BR & 12 (3) & 0.022 & 13 (3) & 0.022 & 12 (3) & 0.013 & 13 (5) & 0.943 & 10.94 \\ 
   & BUD & 4 (17) & 0.006 & 4 (20) & 0.008 & 4 (23) & 0.006 & 37 (20) & \textbf{0.979} & \textbf{6.33} \\ 
   & BAR & 4 (20) & 0.014 & 4 (18) & 0.016 & 4 (23) & 0.014 & 38 (22) & 0.955 & 8.11 \\ 
   & RPW & 11 (4) & 0.014 & 11 (5) & 0.013 & 11 (11) & 0.012 & 17 (4) & 0.960 & 10.03 \\ 
        \cline{2-11}
     70 & BR & 17 (3) & 0.007 & 18 (4) & 0.008 & 17 (4) & 0.003 & 18 (5) & 0.982 & 8.61 \\ 
   & BUD & 5 (26) & 0.002 & 5 (30) & 0.003 & 5 (33) & 0.003 & 55 (30) & \textbf{0.992} & \textbf{4.01} \\ 
   & BAR & 5 (30) & 0.006 & 5 (28) & 0.008 & 4 (33) & 0.008 & 57 (32) & 0.977 & 5.29 \\ 
   & RPW & 15 (4) & 0.005 & 16 (5) & 0.004 & 15 (15) & 0.006 & 24 (7) & 0.984 & 7.59 \\ 

   \hline

 \multicolumn{4}{|c}{\textbf{Scenario 3}} & \multicolumn{7}{l|}{$(\theta_1,\theta_2,\theta_3,\theta_4)=(0.35,0.45,0.7,\textbf{0.8})$}\\

  30 & BR & 7 (3) & 0.012 & 8 (3) & 0.056 & 8 (2) & 0.314 & 8 (4) & 0.618 & 8.12 \\ 
   & BUD & 3 (10) & 0.006 & 3 (8) & 0.024 & 9 (11) & 0.290 & 15 (8) & \textbf{0.679} & \textbf{8.01} \\ 
   & BAR & 2 (12) & 0.012 & 3 (10) & 0.029 & 9 (12) & 0.306 & 15 (11) & 0.652 & 8.99 \\ 
   & RPW & 6 (6) & 0.016 & 6 (3) & 0.034 & 8 (7) & 0.315 & 9 (5) & 0.635 & 8.37 \\ 
        \cline{2-11}
   50 & BR & 12 (3) & 0.005 & 13 (3) & 0.021 & 12 (3) & 0.277 & 13 (5) & 0.697 & 6.57 \\ 
   & BUD & 3 (13) & 0.002 & 4 (11) & 0.006 & 14 (20) & 0.241 & 28 (15) & \textbf{0.751} & \textbf{5.11} \\ 
   & BAR & 3 (22) & 0.003 & 4 (18) & 0.011 & 15 (22) & 0.272 & 28 (19) & 0.714 & 5.86 \\ 
   & RPW & 10 (9) & 0.003 & 11 (2) & 0.014 & 14 (10) & 0.273 & 16 (7) & 0.711 & 6.43 \\ 
      \cline{2-11}
   70 & BR & 17 (3) & 0.001 & 18 (4) & 0.011 & 17 (4) & 0.242 & 18 (5) & 0.747 & 5.57 \\ 
   & BUD & 4 (18) & 0.001 & 5 (17) & 0.001 & 19 (26) & 0.203 & 42 (25) & \textbf{0.795} & \textbf{3.70} \\ 
   & BAR & 3 (32) & 0.001 & 4 (28) & 0.005 & 20 (32) & 0.241 & 42 (28) & 0.752 & 4.61 \\ 
   & RPW & 13 (10) & 0.001 & 15 (2) & 0.003 & 20 (14) & 0.239 & 22 (11) & 0.757 & 5.03 \\  
    \hline

\end{tabular}
\caption{Expected sample size (ESS),   standard deviation (SD),  proportion of simulations that selected  therapy  $a=1, \cdots, 4$  as the most effective treatment ($P_a$) and mean squared error (MSE) 
of the response rate of the best arm for 4-arm trial using 
balanced randomization (BR), 
Bayesian uncertainty directed (BUD) design,  
Bayesian adaptive randomization (BAR) and the randomized play the winner (RPW) design. 
The overall sample size equals either $T=30, 50$ or $70$.
Results are based on 10,000 simulated trials. 
The MSE is scaled by a factor of $10^3$. 
}  \label{TAB_best}
\end{table}

\clearpage
\pagenumbering{gobble}
\begin{table}[ht]
\centering
\footnotesize

     \renewcommand{\arraystretch}{1.1}
\begin{tabular}{| ll  llll | rr |}
\hline
  &    \multicolumn{5}{c}{Expected Sample Size in BMK-positive groups (SD)} & \multicolumn{2}{c|}{Power}\\
\multicolumn{2}{|l}{Biomarker   (BMK)}  &   $1$ & $2$ & $3$ & $4$  & $Po_{+}$ & $Po_{-}$\\
\hline
\bf{Scenario} 1 &   \multicolumn{7}{c|}{arms $a=1, \cdots, 4$ target BMKs $a$,   BMK prevalence (0.5,0.5,0.5,0.5)
} \\
%
 \textbf{BR} & each arm  & 50(7) & 50(7) & 50(7) & 50(7)  &    & \\  
  \cline{2-8}
\textbf{BUD} &control & 91 (11) & 90(11) & 91(11) & 91(11) &   &\\  
 &arm 1 & \textbf{49(30)} & 37(26) & 37(26) & 36(26)  &  10.15 & 9.96\\ 
\hline

 \bf{Scenario} 2 & \multicolumn{7}{c|}{arms $a=1, \cdots, 4$ target BMKs $a$, BMK prevalence (0.5,0.5,0.5,0.5) 
   } 
   \\
\textbf{BR} &control  & 50(7) & 50(7) & 50(7) & 50(7)  &&\\ 
&arm 1 & \textbf{50(7)} & 50(7) & 50(7) & 50(7)  &   {\bf 77.09} & 9.97\\ 
&arm 2 & 50(7) & \textbf{50(7)} & 50(7) & 50(7)  & 9.98 & 9.80 \\ 
  \cline{2-8}
  \textbf{BUD} &control  & 90(12) & 90(11) & 90(10) & 90(10) && \\ 
& arm 1 & \textbf{47(29)} & 38(24) & 38(24) & 38(23) & {\bf 86.92} & 10.00\\
  &arm 2 & 37(26) & \textbf{49(30)} & 37(26) & 38(26) & 10.35 & 10.92\\
\hline

{ \bf Scenario 3} & \multicolumn{7}{c|}{arms $a=1, \cdots, 4$ target BMKs $a$,  BMK prevalence (0.7,0.3,0.5,0.5)
} \\
 \textbf{BR} &control &70(8) & 30(5) & 50(7) & 50(7) && \\ 
& arm 1 & \textbf{70(8)} & 30(5) & 50(7) & 50(7)  &  {\bf86.47} & 10.44 \\ 
&arm 2 &70(8) & \textbf{30(5)} & 50(7) & 50(7)  &  {\bf61.29} & 10.40 \\ 
  &arm 3 & 70(8) & 30(5) & \textbf{50(7)} & 50(7)   &  9.86 & 10.24 \\  
  \cline{2-8}
  
\textbf{BUD} &control & 120(13) & 57(8) & 89(10) & 89(10)  && \\  
&  arm 1 & \textbf{52(33)} & 16(11) & 34(21) & 34(21)  &  {\bf92.52} & 10.81 \\  
& arm 2 & 59(32) & \textbf{35(18)} & 38(22) & 38(22) &  {\bf75.82} & 10.28 \\  
  &arm 3 & 60(38) & 21(16) & \textbf{51(30)} & 39(26)  &  9.87 & 9.44 \\  
 \hline

{\bf Scenario 4}  & \multicolumn{7}{c|}{{arms $1,2,3,4$  target BMKs $1,1,2,3$, BMK prevalence (0.5,0.5,0.5) }} \\
\textbf{BR} &control & 50(7) & 50(7) & 50(7) & & \\ 
&  arm 1 &\textbf{50(7)} & 50(7) & 50(7) &   &  {\bf76.90} & 10.41 \\ 
&arm 2  &\textbf{50(7)} & 50(7) & 50(7) &  &  {\bf 96.15} & 10.53 \\ 
  &arm 3 & 50(7) & 50(7) & \textbf{50(7)}& &  10.08 & 10.61 \\  
  \cline{2-8}
\textbf{BUD} &control & 89(11) & 90(11) & 90(11) &&& \\  
& arm 1 & \textbf{47(27)} & 38(22) & 38(22)&  &  {\bf87.45} & 10.30 \\  
& arm 2 & \textbf{28(18)} & 26(16) & 26(16) &  &  {\bf97.17} & 10.37 \\  
 & arm 3 & 43(30) & \textbf{54(30)} & 41(28) &   & 10.43 & 9.13 \\  
\hline
  
  {\bf Scenario 5} &   \multicolumn{7}{c|}{{  arms 1,2,3,4 target BMKs 1,1,2,2, BMK prevalence (0.5,0.6)}} \\
%
\textbf{BR} & control & 50(7) & 60(7) &   & && \\  
& arm 1 & \textbf{50(7)} & 60(7) & & &  {\bf59.41} & {\bf59.69} \\
&  arm 2 & \textbf{50(7)} & 60(7) & &  &  {\bf76.51} & 10.54 \\  
& arm 3 & 50(7) & \textbf{60(7)} &  & &  {\bf81.93} & 10.07 \\  
&arm 4 & 50(7) & \textbf{60(7)} &  &  &  {\bf97.76} & 10.57 \\  
   \cline{2-8}
\textbf{BUD} & control &85(10) & 102(11) &  &&&\\
&arm 1 & \textbf{38(23)} & 52(27) &  &   &  {\bf61.78} & {\bf67.99} \\ 
&arm 2 & \textbf{52(24)} & 62(30) &  &   &  {\bf89.04} & 9.87 \\ 
&arm 3 & 45(23) & \textbf{54(27)} &  &   &  \bf{92.48} & 9.98 \\ 
&arm 4 & 30(17) & \textbf{29(17)} &  &   & {\bf98.52} & 9.71 \\ 

\hline
\end{tabular}
\caption{{\small Average sample size, standard deviation (SD) and  power in 
the biomarker positive (targeted) and negative subgroups ($Po_+$ and $Po_-$)  for a    5-arm trial with  an overall sample size of $T=500$.
Results are based on 5,000 simulations of a trial using either  BUD or  BR designs. 
For each scenario, the patients' biomarker status has been generated independently for each marker according to the specified prevalences.
 The response rate for the control therapy is constant across scenarios and equal to $0.35$.  
The first three scenarios refer to a trial with four biomarkers, 
and therapy $a=1$ (scenarios 2,3) and  $a=2$ (scenario 3) have  a positive treatment effect (PTE)   for patients with biomarkers $X_{i,a}=1$.  
Scenarios 4 and 5 correspond to a trial with three and two biomarkers respectively.
Therapies $a=1,2$  in Scenario 4 have  PTEs for patients with $X_{i,a}=1$.
In scenario 5 all therapies $a= 2,3,4$ have a PTE in their target population, and arm $a=1$   PTEs  extend  to  all patients.
}}  

\label{TAB::Biomarker}
\end{table}

\end{document}


\maketitle
\clearpage
 
\section{Additional Simulation Results}\label{Additional:Sim}

{\chG  
Consider  two   studies, A and B,  { both}
   with  five  treatment arms and normally distributed  outcomes with variance  $\sigma^2=1$  and  unknown  means $\mu_a$, $a=1, \cdots, 5$.
Study A seeks  to identify   $a^\star = \arg \max_{a} \mu_a$, 
 and attempts to minimize 
the  discrepancy  between   $\mu_{a^\star}$ and $\mu_{\widehat{a}^\star}$,  
where  $\displaystyle \widehat{a}^\star=  \arg \min_{a=1, \cdots, 5} E[ (  \mu_{a} - \mu_{a^\star} ) ^2 | \Sigma_t ]$. 
An  uncertainty metric   consistent with this goal is
  $E[(\mu_{\hat{a}^\star} - \mu_{a^\star} )^2| \Sigma_t ].$ 
In contrast,  the   goal of  study  B is the  estimation of the means  $\mu_a$,  $a=1,\cdots, 5$.
The    goals  are    different and an appropriate  design  for  one   study might  be  inadequate  for the  other.
For study B a balanced     design
  is  consistent  with  the  uncertainty measure 
$\sum_a \mbox{Var}(\mu_a|  \Sigma_t   )$. 
 But the   balanced  design   might be inadequate for  study A.  
   Consider for example a scenario with means $( 1.2,1, 0.2, 0, -0.2)$.
For  study A, in this  case, a    design   that progressively  eliminates  candidate arms  
 performs  considerably  better.   
 With $80$ patients,     
 a  sequential design which  after $10, 20, \dots, 70$ enrollments   
drops   $a$   if   $\displaystyle p({\mu}_a < \max_{1\leq \ell \leq 5} {\mu}_\ell | \Sigma_t ) > 0.95$,  where {\it a priori}   $\mu_a \overset{iid}{\sim}N(0, 1)$,
  selects  $a^\star$
with  a  probability 5\% higher than 
  the  balanced   design.  
}

\subsection*{Simulation Results for Normal Distributed Outcomes}\label{Sec:Normal:Example}
We present  additional  simulation results for a multi-arm trial with a
BUD design as described in Example 1 of Section 3.1 in the main manuscript. 
Outcomes  for patients assigned to arms $a=0, \cdots, K$  are  $N(\theta_a, \sigma^2_a)$ distributed  with known variances  $ \sigma^2_a$ and unknown means $\theta_a \sim N(0, \sigma^2_0)$.
The BUD information measure coincides with the sum of posterior variances $u(\Sigma_t)= -\sum_{k=0}^K Var(\theta_a|\Sigma_t)$. 
Table \ref{Tab:Normal} summarizes selected operating characteristics 
for a trial with $K=3$ experimental arms and $T=150$ patients  in four  scenarios  with $h(t)=3$. 
We consider two scenarios with  identical variances $\sigma^2_a=1$  across arms and two scenarios with different   variances $\sigma^2_a=2,2,1.5, 0.5$ across arms. 
For each scenario Table \ref{Tab:Normal}   shows the asymptotic  approximation of  the patient allocation  (expression \ref{limit} in Section \ref{Sec:Asymptotic}   multiplied by $T=150$). 
Table \ref{Tab:Normal}  indicates that   the average allocation   per arm  agrees closely with the asymptotic  approximation.

\subsection*{Simulation Results for Section \ref{Sec:Multi:Arm1}}
{\chG 
 BUD designs for controlled multi-arm trials can be defined for studies with different aims. For instance one may  consider the minimization of uncertainty  on the treatment effect of (i)  the most effective  experimental arm, 
(ii)   all experimental arms with positive treatment effects or 
(iii) all experimental arms.
We compared BUD designs  with information functions $u_i, i=1,2,3$ ($BUD_i$) defined by distinct study aims,
\begin{align}
u_1(\Sigma_t) &= - \mbox{Var}( \max_a \gamma_{a}    \mid  \Sigma_t ) \label{s1},\\
u_2(\Sigma_t) &= \sum_{a=1}^K \Big[ \mbox{Var}(\gamma_a I(\gamma_a >0) ) - \mbox{Var}( \gamma_a I(\gamma_a>0)   \mid  \Sigma_t ) \Big], \label{s2}\\
u_3(\Sigma_t) &= \sum_{a=1}^K \Big[ \mbox{Var}(\gamma_a)  - \mbox{Var}( \gamma_a   \mid  \Sigma_t ) \Big], \label{s3}
\end{align}
for a trial with two experimental arms $a=1,2$. 
In the four scenarios that we consider,  
the probabilities of response  $(\theta_0, \theta_1, \theta_2)$ 
to arms $a=0,1,2$ equal  $(0.6, 0.5, 0.8)$, $(0.6, 0.5, 0.7), (0.6, 0.4, 0.9)$ and $(0.5, 0.6, 0.7).$ 
The overall sample size $T=3 \times 97$  was selected based on targeted  arm-specific type I and II error bound of $0.05$ and $0.2$ for $(\theta_0,\theta_a)=(0.6, 0.8)$. 

Table \ref{TAB:Inferior:utility} summarizes selected operating characteristics of the three BUD designs. 
Compared to $BUD_3$, a BUD design defined by $u_2$ focuses
 on    positive treatment effects  
and  reduces the average number of assignments to inferior agents  (Scenarios 1 to 3 of Table \ref{TAB:Inferior:utility}) by 58 to 77 patients (34, 37, 20 enrollments on average for  $BUD_2$  compared to 97 under BR). 
A BUD design defined by $u_1$ focuses on the minimization of uncertainty on the treatment effect of the most effective arm, 
and, compared to  $BUD_2$,  
reduces the number of patient assignments to experimental arm one on average  
by 1 to 9 patients in scenarios 1-3.
But when  multiple effective experimental arms exists (scenario 4) 
$BUD_1$  has $10\%$ lower power for arm 1 than $BUD_2$.
}

\clearpage

\section{Additional Figures and Tables}
\subsection{Supplementary Material for Section \ref{Sec:Multi:Arm1}}

\begin{figure}[ht]
\begin{center} 
\includegraphics [width=15cm, height=8cm]  {figures/FigSec3p1Regret}
\caption{
Multi-arm trial, as discussed in Section \ref{Sec:Multi:Arm1} of the manuscript.
Regret function $E_{d^\star }[u(\Sigma_T)] - E_{d}[u(\Sigma_T)]$ between the optimal sequential Bayesian design $d^\star$,
obtained through backwards induction (BI), 
and either  
a Bayesian uncertainty directed design $d=d_{BUD}$ (green curve), 
a balanced randomized design $d=d_{BR}$  (blue  curve), 
Bayesian adaptive randomization $d=d_{BAR}$   (red  curve),
or the randomized play-the winner design $d=d_{RPW}$   (purple curve). 
Results are based on the exact utility $E_{d^\star }[u(\Sigma_T)]$ of $d^\star$ 
under a uniform prior  for the 
outcome probabilities $\theta = (\theta_1, \dots, \theta_4)$.
For $d=d_{BUD}, d_{BAR}, d_{BR}, d_{RPW}$ we computed the expected utility $\mathbb E_{d}[u(\Sigma_T)]$ based on
100,000 simulated trials with  outcome probabilities $\theta = (\theta_1, \dots, \theta_4)$ generated under the prior.  
Exact  computations based on BI  become computationally unfeasible  for  larger  sample  sizes in the order of hundreds of patients.
}
\label{Bandit}
\end{center}
\end{figure}

\clearpage
\input{figures/TabSec3p1Normal}

\clearpage
\input{figures/TabSec3p1Prior}

\clearpage
\input{figures/TabSec3Sensitivity}

\clearpage
\begin{table}[ht]
\centering
\begin{tabular}{|l  r| rrr| rrr|}
  \hline
 BUD                & \multicolumn{1}{c|}{Control} & \multicolumn{3}{c|}{Arm 1}  & \multicolumn{3}{c|}{Arm 2} \\
                 
 design                & ESS (SD) & ESS (SD) & Power & MSE & ESS (SD) & Power & MSE \\ 
  \hline
  
  & \multicolumn{7}{l|}{\textbf{Scenario 1}: $(\theta_0, \theta_1,\theta_2)=(0.6,  {\it 0.5}, {\bf 0.8})$} \\
  $BUD_1$ & 143 (07) & 25 (08) & $\leq1.0$ & 126.1 & 124 (08) & 96.30 & 29.92 \\
  $BUD_2$ & 140 (10) & 34 (21) & $\leq1.0$ & 250.4 & 117 (13) & 97.12 & 28.75 \\
  $BUD_3$ & 121 (04) & 92 (03) & $\leq1.0$ & 050.5 & 077 (05) & 90.09 & 42.57 \\

  \hline 
  & \multicolumn{7}{l|}{\textbf{Scenario 2}: $(\theta_0, \theta_1,\theta_2)=(0.6,  {\it 0.5},  {\bf 0.7} )$}  \\
  $BUD_1$ & 135 (07) & 33 (15) & $\leq1.0$ & 099.6 & 123 (12) & 46.70 & 36.86 \\
  $BUD_2$ & 134 (10) & 37 (22) & $\leq1.0$ & 256.2 & 121 (15) & 47.36 & 39.11 \\
  $BUD_3$ & 118 (03) & 90 (03) & $\leq1.0$ & 053.7 & 084 (04) & 36.91 & 47.27 \\
  \hline 
  & \multicolumn{7}{l|}{\textbf{Scenario 3}: $(\theta_0, \theta_1,\theta_2)=(0.6,  {\it 0.4}, {\bf 0.9} )$}  \\
  $BUD_1$ & 154 (08) & 21 (05) & $\leq1.0$ & 133.7 & 116 (08) & 100 & 22.43 \\
  $BUD_2$ & 161 (11) & 20 (13) & $\leq1.0$ & 254.1 & 110 (10) & 100 & 23.46 \\
  $BUD_3$ & 128 (05) & 97 (04) & $\leq1.0$ & 046.9 & 066 (07) & 99.9 & 34.03 \\

  \hline 
  & \multicolumn{7}{l|}{\textbf{Scenario 4}: $(\theta_0, \theta_1,\theta_2)=(0.5,  {\bf 0.6 },  {\bf 0.7} )$}  \\
  $BUD_1$ & 133 (08) & 50 (25) & 24.1 & 086.7 & 108 (22) & 87.2 & 42.53 \\
  $BUD_2$ & 121 (06) & 78 (17) & 34.8 & 100.6 & 092 (14) & 86.4 & 60.99 \\
  $BUD_3$ & 119 (03) & 88 (03) & 38.6 & 051.1 & 084 (04) & 88.1 & 46.18 \\

   \hline
\end{tabular}

\caption{
Multi-arm trial, as discussed in Section \ref{Sec:Multi:Arm1} of the manuscript.
Sensitivity Analyses using different uncertainty functions to quantify study aims. 
The table shows the expected sample size (ESS), standard deviation (SD), 
power (in \%) for $\mathcal H_{0,a}: \gamma_a \leq 0$ and mean squared error of the estimates arm-specific treatment effect estimates (MSE)
scaled by a factor of $10^3$, for each experimental arm in a controlled 3-arm trial with $336$ patients 
under BUD design $BUD_j, j=1,2,3$ defined by the uncertainty functions (\ref{s1}), (\ref{s2}) and (\ref{s3}).   
}  \label{TAB:Inferior:utility}
\end{table}

\clearpage
\subsection{Supplementary Material for Section \ref{Section:FindBestArm}}

\begin{table}[ht]
\centering
\small
\input{figures/TabSec3p2Sensitivity.tex}
\caption{
Multi-arm trial, as discussed in Section \ref{Section:FindBestArm} of the manuscript.
Expected sample size (ESS),   standard deviation (SD),  proportion of simulations that selected  therapy  a  as the most effective treatment ($P_a$) across 10,000 simulated 4-arm trials with $T=50$ patients using Bayesian uncertainty directed (BUD) sampling using either $h=1,2,6,12$ or $20$ as tuning parameter to define the randomization probabilities.}  \label{TAB:Sensitivity:h}
\end{table}

\clearpage
\subsection{Supplementary Material for Section \ref{Sec:SubgroupTrial}}

\begin{table}[ht]
\centering
\small
\input{figures/TabSec4Sensitivity}
\caption{
Sensitivity  analysis - Multi-arm trial  as presented in Section \ref{Sec:SubgroupTrial} of the article.
 Average sample size, standard deviation (SD) and  power, using Fishers exact test, in 
the biomarker positive (targeted) and negative subgroups ($Po_+$ and $Po_-$)  for a  5-arm trial with  an overall sample size of $T=500$.
Results are based on 5,000 simulations  using either Balanced Randomization or a BUD design with either $w=5, \beta=2,6,8$ (rows 5-16) or  $w=2,5,8, \beta=6$ (rows 17-28). 
The patients' biomarker status has been generated independently for each marker according to the specified prevalences.
 The response rate for the control therapy is constant across scenarios and equal to $0.35$.  
Therapy $a=1$  has  a positive treatment effect (PTE)   for patients with biomarkers $X_{i,a}=1$.}  \label{TABSensitivityBiom}
\end{table}

\clearpage

\begin{table}[ht]
\centering
\small
\input{figures/TabSec4Sensitivity2}
\caption{
Sensitivity  analysis - Multi-arm trial  as presented in Section \ref{Sec:SubgroupTrial} of the article.
Average sample size, standard deviation (SD) and  power, using Fishers exact test, in 
the biomarker positive (targeted) and negative subgroups ($Po_+$ and $Po_-$)  for a  5-arm trial with  an overall sample size of $T=500$.
Results are based on 5,000 simulations using a BUD design with either $w=5, \beta=2,6,8$ (rows 6-20) or  $w=2,5,8, \beta=6$ (rows 21-35). 
The patients' biomarker status has been generated independently for each marker according to the specified prevalences.
 The response rate for the control therapy is constant across scenarios and equal to $0.35$.  
Scenarios 4 and 5 correspond to a trial with three  biomarkers  and 
therapies $a=1,2$  have  PTEs for patients with $X_{i,a}=1$.
}  \label{TABSensitivityBiom2}
\end{table}

\clearpage

\begin{table}[ht]
\centering
\small
\input{figures/newTableBinaryMultiplenew.tex}
\caption{
Multi-arm trial  as presented in Section \ref{Sec:CoPrimary} of the manuscript.
Power and mean squared error (scaled by a factor of $10^3$) for each experimental arm in a controlled five-arm trial with  an overall sample size of $T=348$ patients. 
Results are based on 5,000 simulated trials using either balanced randomization (BR), 
a BUD design  for testing treatment effects
($\text{BUD}_1$), 
and a BUD design  for estimating treatment effects 
($\text{BUD}_2$).
The outcome probabilities for the control arm and experimental arms $a=1,2,3$  equal  $\theta_a = (0.15,0.25,0.4,0.2)$ for the event of a positive response on both endpoints (0.15), for endpoint 1 (0.25) or endpoint 2 (0.4) only, 
and no positive response (0.2).
The targeted type I error rate  is  set at $\alpha =0.05$.
}  \label{TABmulti}
\end{table}

\clearpage

\section{Proofs}

\begin{proof}[Proof of Lemma \ref{asymptotics::lemma}] 
We show that for any $\varepsilon>0$, the process $ \widehat p_{0,t}$ can only visit the interval $[\rho_0+\varepsilon,1]$ finitely often. A symmetric argument can be applied to intervals lying below $\rho_0$.

Define the martingale
\eq{
M_{t} =  \widehat p_{0,t} - \sum_{i=1}^t \E[\widehat p_{0,i}-\widehat p_{0,i-1}\mid \Sigma_{i-1}] .
}
As $M_{t+1}-M_{t}\leq \mathcal O(1/t)$, Doob's $L^2$ martingale convergence theorem implies that $M_{t}\to M_{\infty}<\infty$ a.s. Furthermore, a simple derivation yields
\eq{
\E[\widehat p_{0,t+1}- \widehat p_{0,t}\mid \Sigma_t]= 
-\frac{ \widehat p_{0,t}}{t+1} + \frac{1}{t+1}\frac{\Delta_{t}^h(0)}{\Delta_{t}^h(0)+\Delta_{t}^h(1)} = \frac{F_t}{t+1}.
}
In order to prove that $ \widehat p_{0,t}=M_{t}+ \sum_{i=1}^t \E[\widehat p_{0,i}-\widehat p_{0,i-1}\mid \Sigma_{i-1}] $ only visits $[\rho_0+\varepsilon,1]$ finitely often, we will use the convergence of the first term, and the fact that the summands in the second term define a drift toward $\rho_0$ for large values of $i$.

On a set of probability 1, there is a number $N$ such that for any $t>N$, $| M_{t}-M_{\infty}|<\varepsilon/4$, and $F_t<-c<0$ when $ \widehat p_{0,t}\in[\rho_0+\varepsilon/2,1]$. Therefore, the trajectory $(\widehat p_{0,N+k})_{k\geq 1}$ cannot increase more than $\varepsilon/4$ while staying in $[\rho_0+\varepsilon/2,1]$. As $\sum_{t=1}^\infty(t+1)^{-1}=\infty$, every time the process enters this interval, it will eventually exit below $\rho+\varepsilon/2$. Finally, observe that $|F_t|\leq 2$ and thus $|F_t/(t+1)|\to 0$, so for $t$ large enough, the process cannot reenter the interval $[\rho+\varepsilon/2,1]$ above $\rho+3\varepsilon/4$, so it can only enter $[\rho+\varepsilon,1]$ finitely often.
\end{proof}

\begin{proof}[Asymptotic allocation for binary outcomes model.] Due to the conjugacy between the Bernoulli likelihood and the Beta prior, the variables $\theta_0,\dots,\theta_K$ are independent and Beta a posteriori. This makes it possible to write down the information gain in closed form. Letting $\hat\theta_{a,t}$ be the proportion of successes in arm $a$ by time $t$, 
\eq{
\Delta_{t}(a) =\;& -\frac{(\alpha+t  \widehat p_{a,t}\hat\theta_{a,t}+1)(\beta+t  \widehat p_{a,t}(1-\hat\theta_{a,t}))}{(\alpha+\beta+t  \widehat p_{a,t}+1)^2(\alpha+\beta+t  \widehat p_{a,t}+2)} 
\left( \frac{t \widehat p_{a,t}\hat\theta_a+\alpha}{t \widehat p_{a,t}+\alpha+\beta} \right)\\
& - \frac{(\alpha+t  \widehat p_{a,t}\hat\theta_{a,t})(\beta+t  \widehat p_{a,t}(1-\hat\theta_{a,t})+1)}{(\alpha+\beta+t  \widehat p_{a,t}+1)^2(\alpha+\beta+t  \widehat p_{a,t}+2)} 
\left( \frac{t \widehat p_{a,t}(1-\hat\theta_a)+\beta}{t \widehat p_{a,t}+\alpha+\beta} \right)\\
& +\frac{(\alpha+t  \widehat p_{a,t}\hat\theta_{a,t})(\beta+t  \widehat p_{a,t}(1-\hat\theta_{a,t}))}{(\alpha+\beta+t  \widehat p_{a,t})^2(\alpha+\beta+t  \widehat p_{a,t}+1)} \\
=\;& \frac{1}{t^2}\left(\frac{\hat\theta_{a,t}(1-\hat\theta_{a,t})}{ \widehat p_{a,t}^2}\right) + \mathcal O((t\widehat p_{a,t})^{-3}). \\
=\;& \frac{1}{t^2}\left(\frac{\theta_{a,t}(1-\theta_{a,t})}{ \widehat p_{a,t}^2}\right) +\mathcal O(|\hat\theta_{a,t}-\theta_a|(t\widehat p_{a,t})^{-2})+ \mathcal O((t\widehat p_{a,t})^{-3}).
}
where the last two equalities hold due to Taylor's theorem. 

Now, we can check the condition of Lemma \ref{asymptotics::lemma} by the same argument used in the normal outcomes case. Writing $\sigma_a=\theta_a(1-\theta_a)$, the Taylor series approximation implies
\eq{
F_t = - \widehat p_{0,t}+\frac{ \left(\widehat p_{0,t}^{-2}\sigma_0^{2}
+\mathcal O(|\hat\theta_{0,t}-\theta_0|\widehat p_{0,t}^{-2})+ \mathcal O(t^{-1}\widehat p_{0,t}^{-3})\right)^h}
{\left(\widehat p_{0,t}^{-2}\sigma_0^{2}
+\mathcal O(|\hat\theta_{0,t}-\theta_0|\widehat p_{0,t}^{-2})+ \mathcal O(t^{-1}\widehat p_{0,t}^{-3})\right)^h
+ \left(\widehat p_{1,t}^{-2}\sigma_1^{2}
+\mathcal O(|\hat\theta_{1,t}-\theta_1|\widehat p_{1,t}^{-2})+ \mathcal O(t^{-1}\widehat p_{1,t}^{-3})\right)^h }.
}
We show that the residual terms in $\mathcal O(\cdot)$ are negligible in the limit, which is trivial when $\widehat p_{0,t}$ and $\widehat p_{1,t}$ are of constant order. For any $\delta\in(0,1/4)$, on the subsequence where $\widehat p_{a,t}> \max\{(t\widehat p_{a,t})^{-1/4+\delta} \allowbreak (\log\log (t\widehat p_{a,t}))^{1/2}, \allowbreak t^{-1/3+\delta}\}$ for $a=0,1$, by the law of the iterated logarithm, the residual terms in the above expression go to 0 a.s. as $t\to\infty$. Otherwise, one of $\widehat p_{0,t}$ or $\widehat p_{1,t}$ goes to $0$ as $t\to\infty$. We conclude that, on a set of probability 1, $F_t$ converges to 
\eq{
\tilde F_t = - \widehat p_{0,t}+\frac{ \widehat p_{0,t}^{-2h}\sigma_0^{2h}}
{\widehat p_{0,t}^{-2h}\sigma_0^{2h}
+ (1-\widehat p_{0,t})^{-2h}\sigma_1^{2h}.}
} 
The argument used in the normal model allows us to apply Lemma \ref{asymptotics::lemma} and Remark \ref{multiarm::remark}, which imply the a.s. convergence $\widehat p_{a,t}\to\rho_a\propto \sigma_a^{2h/(2h+1)}$ for a BUD design with $K$ arms.

\end{proof}

\section{ Computational Details }\label{Sec:Computational:Details}

The  computing time to simulate a BUD trial increases linearly 
with respect to the sample size $T$ and with respect to the number of  possible actions.
In our examples the computing time for  BUD simulations, as expected, is very similar to BAR, 
and   significantly faster  than  computing  the optimal Bayesian design    with   dynamic programming.

\noindent {\bf Computational details for Section \ref{Exa1}: } 
In the first multi-arm trial designs of Section \ref{Exa1}, the  utility  
$
u(\Sigma_t) = \sum_{a=1}^K ( v_a - \Var(\gamma_a | \Sigma _t)  )
$,
with independent and conjugate priors Beta$(1, 1 )$ for the response rates  of each arm, has a closed form.
The expected utility conditional on $A_{t+1}=a$ is computed with the formula 
$E[ u(\Sigma_{t+1}) \mid \Sigma_t, A_{t+1}=a ] = 
u_{a,1}\widehat \theta _a(t) + 
u_{a,0}(1- \widehat \theta _a(t) ),$
where $\widehat \theta _a (t) = E[\theta_a \mid \Sigma _t]$, and $u_{a,0}$ and $u_{a,1}$ are the utilities at time $t+1$ with a negative and positive outcome, respectively. 
In the second example 
we computed the
negative entropy $u(\Sigma_t) = E[ \log p_{\theta_{a^\star}} (\theta_{a^\star} \mid \Sigma_t)\mid \Sigma_t ]$ 
for the posterior of the most effective arm $a^\star$ using numerical integration.

\noindent {\bf Computational details for Section \ref{Sec:SubgroupTrial}: } 
The prior specified on the response rates $\theta$ makes it possible to compute the posterior probability of treatment effects $E_a=(E_{0,a},E_{1,a})$ in arm $a$ in closed form, which yields an exact expression for the information gain $\Delta_a(t)$. For simplicity, consider the case with one control and one experimental arm, $a=0,1$. Here, 
\eq{
p(E_1\mid Y_{1:t}) = \int p(Y_{1:t}\mid \theta) p(\theta\mid E_1) p(E_1) d\theta
}
where $\theta= (\theta_{0,0},\theta_{1,0},\theta_{0,1},\theta_{1,1})$. Inside the integral, $p(E_1)$ is a simple function of $\pi$ and $\lambda$, while $p(\theta \mid E_1)$ is a product Beta density with integer parameters restricted to a set $R\subset [0,1]^{4}$. The normalizing constant of this truncated density can be computed exactly using a well-known combinatorial formula, derived by representing Beta random variables as order statistics of uniform random variables. By conjugacy, the product $p(Y_{1:t}\mid \theta)p(\theta\mid E_1)$ is proportional to a product of beta densities, and its integral over $R$ is obtained using the combinatorial formula mentioned.

For each treatment $a=1, \dots, K$ we tested treatment effects in the biomarker-targeted population 
(all patients with biomarker characteristic $X_{t,b_a}=1$) and in the  biomarker-negative subgroup 
(patients with $X_{t,b_a}=0$) using a parametric bootstrap test (see \cite{ventz2017combining} for details).

\noindent {\bf Computational details for Section \ref{Sec:CoPrimary}: } 
In the BUD design
for multi-arm trials with two co-primary endpoints, 
we estimated  the posterior probability of treatment effects by Monte Carlo integration, drawing independent samples from the Dirichlet posterior of the response rates $\theta$. 
The expected utility is then computed by 
$E[ u(\Sigma_{t+1}) \mid \Sigma_t, A_{t+1}=a ] =  \sum_{y \in \{0,1\}^2} u_{y, a}(t+1) \times \widehat \theta _{y, a}(t),$ 
where  $\widehat \theta _{y, a}(t) = E[ \theta _{y, a}\mid  \Sigma_t ]$
and $u_{y, a}(t+1)$ indicates the expected utility  $u(\Sigma_{t+1})$ given $\Sigma_t$ and $(A_{t+1}, Y_{t+1})=(a,y)$.

\bibliographystyle{unsrt}
\bibliography{Bibliography}
\addcontentsline{toc}{chapter}{Bibliography}